%% file: 3rd_for_postdoc.tex
\documentclass[twocolumn]{aastex62}
\pdfoutput=1 
 
\def\secref#1{Section~\ref{#1}}
\def\figref#1{Figure~\ref{#1}}
\newcommand{\change}[1]{#1}
\newcommand{\mathchange}[1]{#1}
\newcommand{\changetwo}[1]{#1}
\newcommand{\changethree}[1]{#1}

\begin{document}

\title{Multi-spacecraft observations of coronal loops to verify a force-free field reconstruction and infer loop cross sections}
\author{Marika I. McCarthy}
\affiliation{Montana State University}

\author{Dana W. Longcope}
\affiliation{Montana State University}

\author{Anna Malanushenko}
\affiliation{Montana State University}
\affiliation{HAO}

\begin{abstract}
Active region EUV loops are believed to trace a subset of magnetic field lines through the corona.  \cite{Malanushenko2009} proposed a method, using loop images and line-of-sight photospheric magnetograms, to infer the three-dimensional shape and field strength along each loop. \citet{McCarthy2019} used this novel method to compute the total magnetic flux interconnecting a pair of active regions observed by SDO/AIA. They adopted the common assumption that each loop had a circular cross section. The accuracy of inferred shape and circularity of cross sections can both be tested using observations of the same loops from additional vantage points as provided by STEREO/EUVI. Here, we use multiple viewing angles to confirm the three-dimensional structure of loops.  Of 151 viable cases, 105 (69.5\%) matched some form of visible coronal structure when viewed approximately in quadrature. A loop with a circular cross-section should appear of a similar width in different perspectives. In contradiction to this, we find a puzzling lack of correlation between loop diameters seen from different perspectives, even an anti-correlation in some cases. Features identified as monolithic loops in AIA may, in fact, be more complex density enhancements. The 30.5\% of reconstructions from AIA which did not match any feature in EUVI might be such enhancements. Others may be genuine loop structures, but with elliptical cross sections. We observe an anti-correlation between diameter and brightness, lending support to the latter hypothesis. \changetwo{Of 13 suitable for width analysis, f}our loops are consistent with non-circular cross sections, where we find anti-correlation in both comparisons.

\end{abstract}

\section{Introduction}

When viewed in extreme-ultraviolet (EUV) or X-ray wavelengths, the Sun's atmosphere is composed of narrow strands called coronal loops.  Their enhanced density compared to the background causes these coronal loops to emit brightly in these wavelengths.  The prevailing view is that the coronal plasma, aligned to the magnetic field, creates a cylindrical flux tube that is isotropic in the direction perpendicular to the field line \citep{2014LRSP...11....4R}.  \change{With each new generation of EUV \changetwo{and soft X-ray} space telescope\changetwo{s} \citep[e.g.,  SXT, TRACE, Hi-C, and others;][]{1991SoPh..136...37T, Handy1999, 2014SoPh..289.4393K} the observation of these loops and literature reporting their physical properties grows.}  
However, the idea that these observed loops are monolithic objects, and that the cross sections of the flux tube\change{s} are indeed circular, have been called into question \citep{2013ApJ...775..120M}.

Direct measurements of magnetic fields throughout the coronal volume are \change{extremely difficult}.  Thus, one of the main methods to infer the coronal magnetic field is the traditional non-linear force-free field (NLFFF) extrapolation method using vector field measurements taken at the solar surface \citep{DeRosa_2009,2012LRSP....9....5W}.  A volume-filling field is computed from the photospheric magnetic boundary conditions under the assumption that the magnetic force vanishes throughout.  Alternatively, under the assumption that the loops are bundles of field lines that have been energized and made visible, mere observation of coronal loops provides insight into the coronal field.

%\begin{itemize}
%\item traditional nlfff extrapolation
%\item direct measurements of coronal fields trhoughout the coronal volume are impossible so they are inferred in several ways
%\item nofff vectorfield measurements at the photosphere
%\end{itemize}

The method of \cite{Malanushenko2009}, however, uses both extrapolation and observation to determine the field structure in the corona.  This method varies the value of parameters $\alpha$ (twist) and $h$ (height above the solar surface along \change{the line of sight}) to fit each coronal loop to its own linear force-free field (LFFF), using boundary conditions from a line-of-sight (LOS) photospheric magnetogram and tracing a portion of the loop in an EUV image.  We will hereafter call this method $\alpha$-h fitting.  The loops are matched in the plane-of-sky (POS) by design, and this method yields a three-dimensional structure from these two-dimensional observations. 

Following its development, the $\alpha$-h fitting has been applied in several instances \citep{2011ApJ...736...97M, 2012ApJ...756..153M, 2014ApJ...783..102M,2015SoPh..290..491V, McCarthy2019} but has been exclusively used from one viewing perspective.  By design, it is in that perspective which the model was fit.
%\begin{enumerate}
%\item following its development we’ve applied this in several cases but has not really been 
%\item only been used in one POS perspective, and by design it is the one that is fit
%\item it is our objective  / one of the aims of this work is to provide a / is to test its 3d construction from a second perspective different perspective
%\item 
%\end{enumerate}
While this approach is novel in its combination of types of data it synthesizes to determine the coronal field, it has not been further tested.  The way this work will test the $\alpha$-h fitting is to get a second perspective in order to verify the validity of the 3D structure.  The twin Solar TErrestrial RElations Observatory (STEREO) spacecraft \citep{2008SSRv..136....5K} and the Extreme UltraViolet Imager \citep[EUVI,][]{2004SPIE.5171..111W} aboard it, provide the second perspective required for elucidating the elusive third dimension.  

The STEREO mission has a fruitful history of using the twin spacecraft to do stereoscopic %and tomographic 
analysis \citep[see][for a review]{2011LRSP....8....5A}.  Typical usage of STEREO for such analysis first identifies a point of interest in the image from one instrument.  This feature lies along some LOS from the first instrument.  This LOS is then projected into the POS from the second viewing angle, and features along this line are identified as candidates corresponding to the point of interest in the first image.  

This conventional usage of STEREO for \change{stereoscopy} relies, however, on the feature even existing (i.e., being observable) from this second viewing angle.  If the feature does not exist (i.e., is not observed from the second perspective), it is possible there is a false identification of correspondence between the features from both perspectives.  This ties into a selection bias posited by \citet{2013ApJ...775..120M}: potential asymmetries in the diameters of the loops lead to differing amounts of emitting loop plasma (and thus, brightness) along these varying lines of sight.  In such a case, something appearing as a loop in one perspective would not even appear in another.

With the aid of the $\alpha$-h fitting, we need not rely on the second perspective to construct the 3D information of the loop in which we are interested.  The structure is extrapolated from only a single perspective.  It is now possible for a point along a loop reconstruction to be unambiguously mapped into the field of view (FOV) of the other instrument, without the need for triangulation.

In addition to verifying the loop reconstruction from the $\alpha$-h fits, we can use the observation from multiple spacecraft to infer properties of loops.  Specifically, we can add to the growing literature of \change{investigations into loop asymmetries} % \change{for (or against)} loop asymmetries %\change{observations to investigate} loop asymmetries %\colorbox{cyan}{observational evidence of loop asymmetries} 
as proposed by \citet{2013ApJ...775..120M} \change{using direct observations of loops} \citep{Kucera2019,2020ApJ...900..167K}.

%\change{In previous work on coronal loops and their cross sections from various instruments...}
%\change{\colorbox{cyan}{From referee's report:}  Include prev work on coronal loops and cross sections from SXT, TRACE, EUVI, AIA, Hi-C}

\cite{McCarthy2019} presented a case study, which provides an ideal starting point for this investigation. The data set contained 301 loops which were cataloged between emerging active region NOAA AR11149 adjacent to AR11147 for approximately 48 hours beginning 2011 January 20 at 22:01UTC.  A subset of these were fit at 56 instances in time using this $\alpha$-h-fitting method.  The EUV data was obtained from the $171\AA$ channel of the Atmospheric Imaging Assembly \citep[AIA,][]{2012SoPh..275...17L} with LOS magnetograms provided by the Helioseismic and Magnetic Imager \citep[HMI,][]{Scherrer2012}, both aboard the Solar Dynamics Observatory (SDO).  

The LFFF models from this paper's data set have already proved useful in other instances, like in investigating heating from localized reconnection at topological boundaries \citep{Longcope2020}.  Further, the results of \cite{McCarthy2019} overcounted the flux interconnecting the two ARs, which may be attributed to the assumption in their analysis that coronal loops had circular cross sections.  

This work will use the additional vantage points provided by STEREO/EUVI to achieve two related goals: first to assess the goodness of fit of the $\alpha$-h fitting from a different instrument's LOS, and then to use the multiple viewing angles to analyze loop diameters from different planes of the sky.

This investigation will be detailed in the following sections.  We discuss the acquisition of the EUVI data sets that compliment that used in \cite{McCarthy2019} in \secref{sec:dataacq} and alignment of the $\alpha$-h fits with the new data set in \secref{sec:align}.  Due to limitations of the data, some analysis is best performed by a qualitative, visual inspection.  That analysis is detailed in \secref{sec:byeye}.  A quantitative analysis is undertaken in \secref{sec:quant}.  The final section details conclusions drawn from our results and future outlook in \secref{sec:conclusions}.

\section{Data set acquisition}\label{sec:dataacq}

\begin{figure}[tbph]
\centering
     \includegraphics[width=0.999\linewidth]{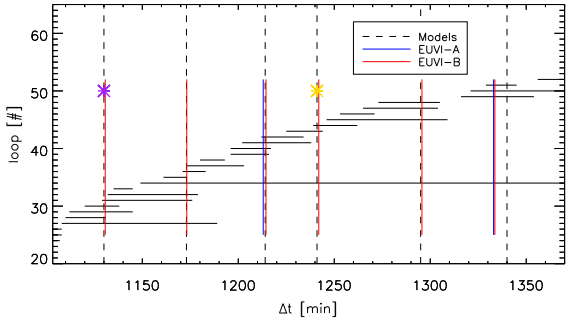}
    \caption{A subset of the data is shown to illustrate how the loops, $\alpha$-h fitting, and STEREO data are related.  Horizontal lines are the loops and their durations.  Vertical dashed line is the instance the $\alpha$-h fitting was done.  The colored vertical lines are the data used from EUVI.  Spacecraft A data is in blue and B is in red.  The purple and yellow stars correspond to the instances shown in Figures \ref{fig:example2} and \ref{fig:example3}, with 5 and 3 loops respectively (corresponding to the intersections with horizontal bars).  As the original data set was constructed from SDO/AIA images, each time step (in minutes) has a corresponding AIA image.}
    \label{fig:times}
  \end{figure}
  
  \begin{figure*}[htbp]
\centering
     \includegraphics[width=0.99\linewidth]{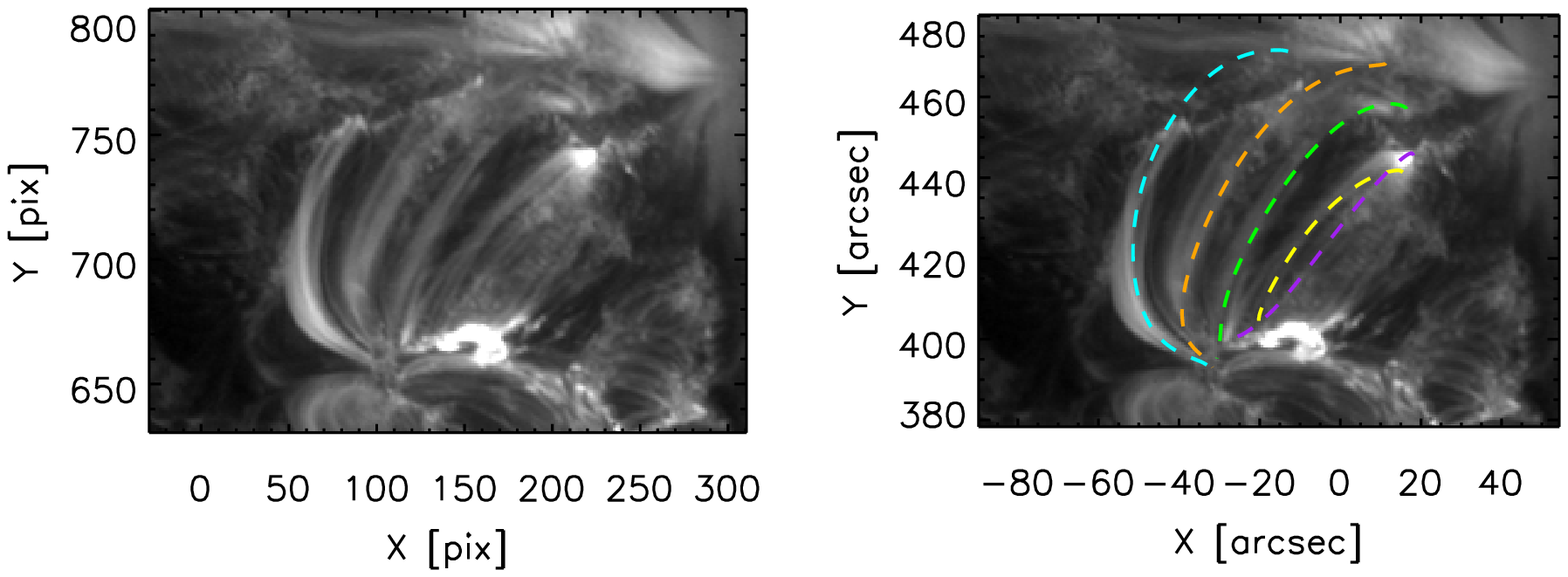}
    \caption{An image from AIA $\mathchange{171}$\change{\AA} with the $\alpha$-h fits overplotted.  This corresponds to the time of the purple star and its $5$ loops in \figref{fig:times}.  This figure has been square-root-scaled with color table saturating at $70 \sqrt{\text{DN/s}}$ for clarity.  Later, we denote this -- and specifically the fit plotted in purple --  as Example 1 (see \figref{fig:example2}).}
    \label{AIAex}
  \end{figure*}

The $\alpha$-h fits presented in \cite{McCarthy2019} were constructed at 56 different time instances from 185 individual loops out of the 301 that were identified interconnecting the ARs. This is illustrated in \figref{fig:times}, which shows a four hour subset of the total 48-hour interval.  Horizontal lines show the cataloged loops (with their extent representing their durations) and the dashed vertical lines are the times that the $\alpha$-h fitting was performed.  The loops that were fit during this process are intersected by the dashed vertical lines.  The purple star marks a time whose image, shown in \figref{AIAex}, includes five loops, shown as colored curves on the right panel.  These loops correspond to the five horizontal lines crossing that one vertical dashed line on \figref{fig:times}.  As the modeling times were chosen to maximize the number of loops fit, some loops were modeled at multiple times (e.g., the long-lasting loop \#34 in \figref{fig:times}).  Thus the data set contains 199 different $\alpha$-h fits representing 185 distinct loops.

The EUVI data was obtained from the Virtual Solar Observatory \citep[VSO,][]{VSO} using SolarSoft \citep{FreelandHandy1998}.  \change{The angular separation from Earth was 86$^\circ$ and 92$^\circ$ for EUVI-A and EUVI-B, respectively, with separation angle of 178$^\circ$ between them.}  The $171\AA$ passband was chosen as the $\alpha$-h fitting used the same passband in AIA.  For each of the discrete times at which the $\alpha$-h fitting was done, we looked within an half hour interval on either side of the desired time and sought the image taken nearest to that time.  In \figref{fig:times}, the time of the EUVI-A data is denoted by a blue vertical bar and the EUVI-B data is in red.  The vast majority of the EUVI images used are within 10 minutes of the corresponding model's time.  This is within the median lifetime of these loops \citep{McCarthy2019}.  These data were processed using the SolarSoft \citep{FreelandHandy1998}  {\tt secchi\_prep} command with keyword {\tt /dn2p\_off} to preserve intensity units as DN/s.  Some images were discarded for issues with data quality, primarily blank images.  In total the data set contains 18 images from STEREO-A and 26 from STEREO-B, with 17 instances of both A and B observing simultaneously.

The SDO/AIA data used to compare with the EUVI images was previously downloaded through VSO via SolarSoft and prepped using {\tt aia\_prep} with exposure normalization.  It is the same data  set used in \cite{McCarthy2019} to perform the $\alpha$-h fitting.  \figref{AIAex} shows an instance (denoted by the time of the purple star in \figref{fig:times}) of the $\alpha$-h fits plotted over top of the companion AIA image.

\section{Alignment of model with data}\label{sec:align}

The $\alpha$-h fitting method of \cite{Malanushenko2009} builds a set of volume-filling LFFFs from a photospheric magnetogram, and then uses a manually-traced portion of an observed coronal loop from an EUV image to determine the LFFF to which it fits.  An LFFF, or a constant-$\alpha$ field, is a particular case of a force-free field (FFF),
\begin{equation}\label{fff}
\nabla \times \bold{B} = \alpha(\bold{r})\bold{B},
\end{equation}
in which $\alpha$ is uniform in space, $\nabla \alpha = 0$, and thus serves as a parameter of the model.  %The magnetic fields in the corona are believed to be in a FFF satisfied by that equation.  
This condition on $\alpha$ and the condition $\nabla \cdot \bold{B} = 0$ allow us to transform Equation \ref{fff} into a Helmholtz equation for $\bold{B}$.  The volume-filling fields were constructed in the half space ($z \geq 0$) using the tangent plane approximation (i.e., in rectilinear box tangent to a point on the solar surface) by solving the Helmholtz equation for $\bold{B}$ \citep{1977ApJ...212..873C, Lothian1995}.  An HMI LOS magnetogram \citep{2012SoPh..275..229S} provided the boundary conditions used in the creation of these fields.  

For this data set, LFFFs for 61 values of $\alpha$ were computed, with $\alpha$ equally spaced in the range [-0.05, 0.05] arcsec$^{-1}$.  To link the information given by the photospheric extrapolation to that from the EUV data, a coronal loop is manually traced with a smooth curve.  (Note that for this method, we need only trace a portion of the loop.  The feet of the loop need not be included.)  The parameters $\alpha$ and $h$ (height, distance along the line of sight of a point at the center of the loop) are varied, field lines are extrapolated from the POS location of midpoint of the trace for various values of $h$, and the extrapolated field line POS projections are compared to the trace.  An average distance function between the extrapolation and the trace, $d(h,\alpha)$, is minimized to choose the correct model for that field line.  The $\alpha$-h fitting yields a 3D model of the LFFF line to which the observed AIA loop was fit, which we will hereafter call a ``track.''  As these tracks are three-dimensional, they can be rotated such that they can be observed from both AIA (by construction, see \figref{AIAex}) and STEREO (through a coordinate transformation).

\begin{figure*}[htbp]
\centering
\hspace{0.4 in}
     \includegraphics[width=0.46\linewidth]{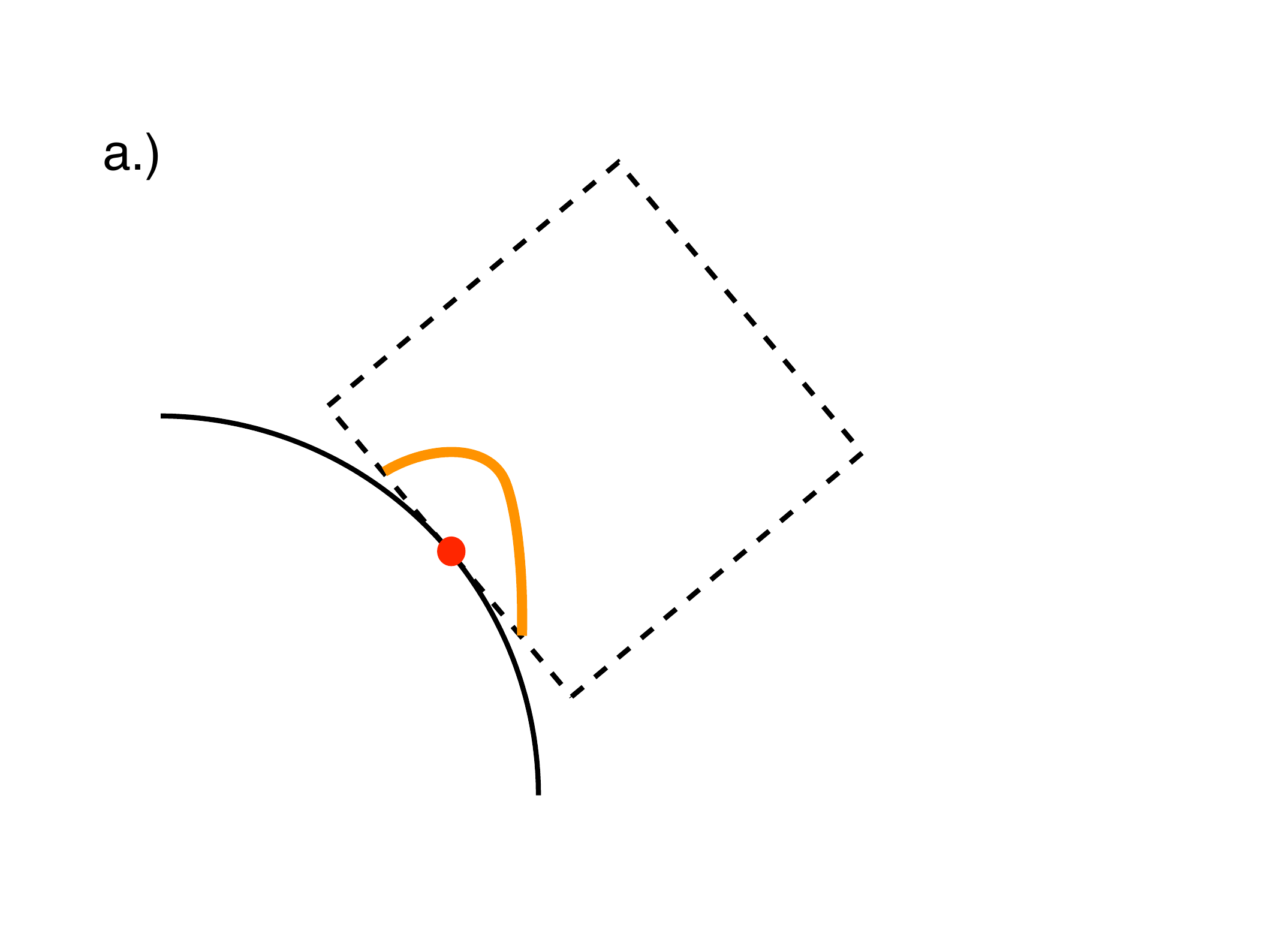}
     \includegraphics[width=0.46\linewidth]{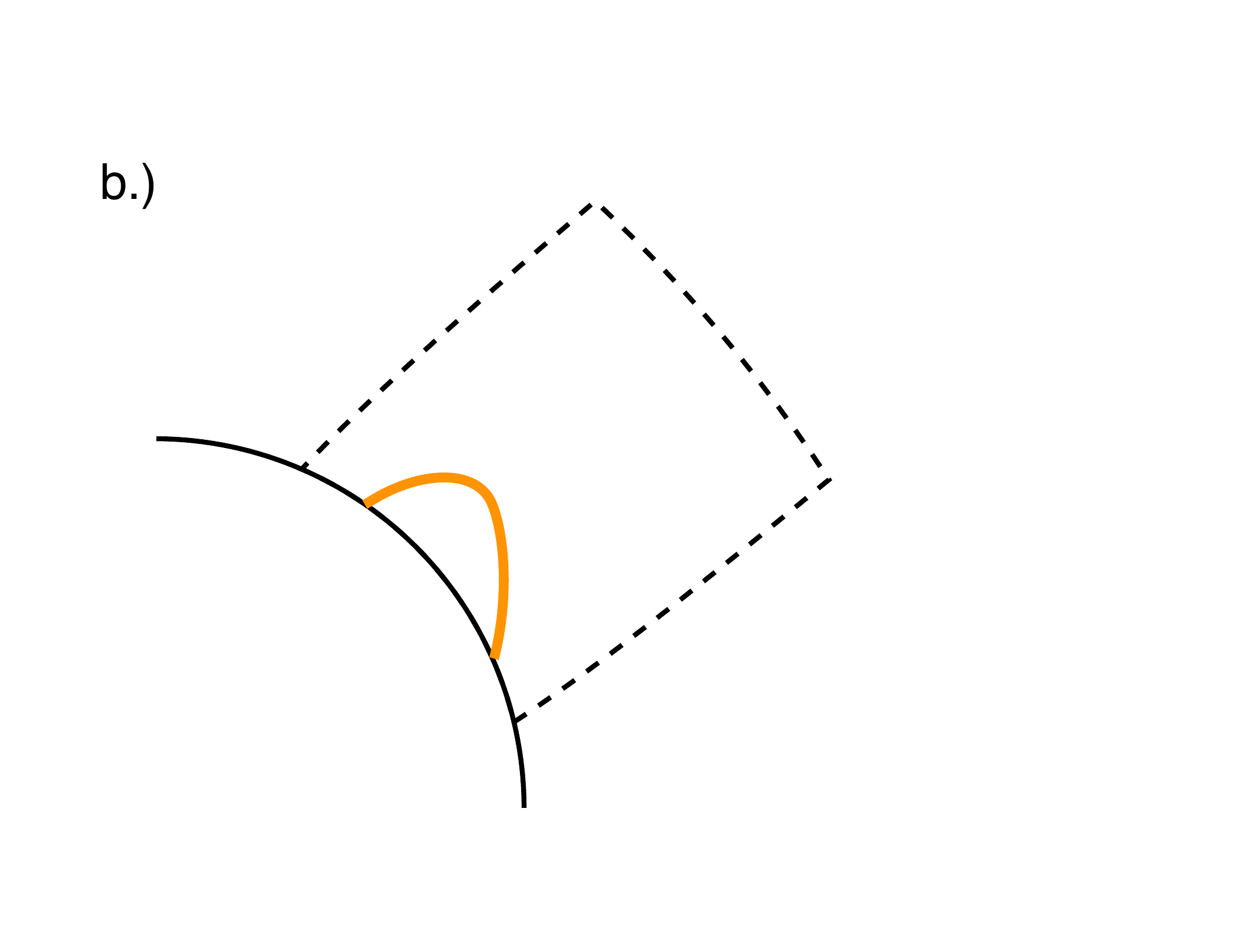}
    \caption{Panel (a) shows the rectilinear box in which the $\alpha$-h fits are constructed.  The point of tangency (red circle) is the only location that lies on the solar surface.  Panel (b) shows the approximate shape of the box after the deformation is applied.  The bottom surface of this volume now entirely lies on the solar surface, not just only at the point of tangency.  In both panels, a track constructed from the $\alpha$-h fitting represents a coronal loop and is depicted in orange.}
    \label{fig:deformation}
  \end{figure*}

As an independent verification of the $\alpha$-h fitting method we now assess how well the tracks it produces match coronal loops seen from different viewpoints offered by STEREO.  The track (i.e., the three-dimensional curve produced from an AIA image by $\alpha$-h fitting) is projected onto the image plane of STEREO and plotted over an EUVI image.  The tangent-plane approximation used in the modeling poses one difficulty in making this comparison.  If the rectilinear box in which the tracks were constructed was directly transformed, then this box would still lie on a plane tangent to a point on the solar surface (see panel (a) in \figref{fig:deformation}).  While near the point of tangency this was a good approximation, further away in the North-South direction we end up with the tangent plane's lower surface deviating from the true solar surface -- that is, it should be on the limb.

%\footnote{\colorbox{cyan}{rotated about Sun's center by theta and phi of spherical coordinates} (by the way, since you describe the other transformation below in detail, why not say a sentence about the rotations here? or a footnote. E.g., 'first rotate by lon to put the AR to the central meridian, then by -lat to put it at the equator, then by lon1 to put to the new longitude and by lat1 to put to the new latitude') or smth like this}

To remedy the discrepancy in the location of the solar surface from the tangent plane, we apply a deformation of the rectilinear box, and therefore the track, before rotating into the new EUVI LOS.  This transformation is described as follows:  in the tangent plane, each point in a track is defined by the three Cartesian coordinates $(x, y, z)$ with equivalent spherical coordinates $(r, \theta,\phi)$.  In the tangent plane model, the $z$-coordinate is the height above the solar surface ($z=0$).  However, only the point of tangency actually lies on the solar surface; the other points in the $z=0$ plane lie above, $r > R_{\odot}$. We wish to define $r$ coordinate, $r_{\text{new}}$, such that this new $r$-coordinate gives the radial distance including height above the solar surface, $R_\odot$.  The $\theta$ and $\phi$ coordinates remain unchanged while we do this adjustment.  Thus we define $r_{\text{new}}  = R_\odot + z$.  We then transform $(r_{\text{new}}, \theta, \phi)$ back into Cartesian coordinates $(x',y',z')$ to be used subsequently.  The resultant deformation is \change{approximately} shown in panel (b) of \figref{fig:deformation}, where the lower surface of the modeling volume (and the footpoints of the track) lies entirely on the solar surface.

\section{Qualitative comparison of tracks to EUVI images}\label{sec:byeye}

\begin{figure}[htbp]
\centering
     \includegraphics[width=0.999\linewidth]{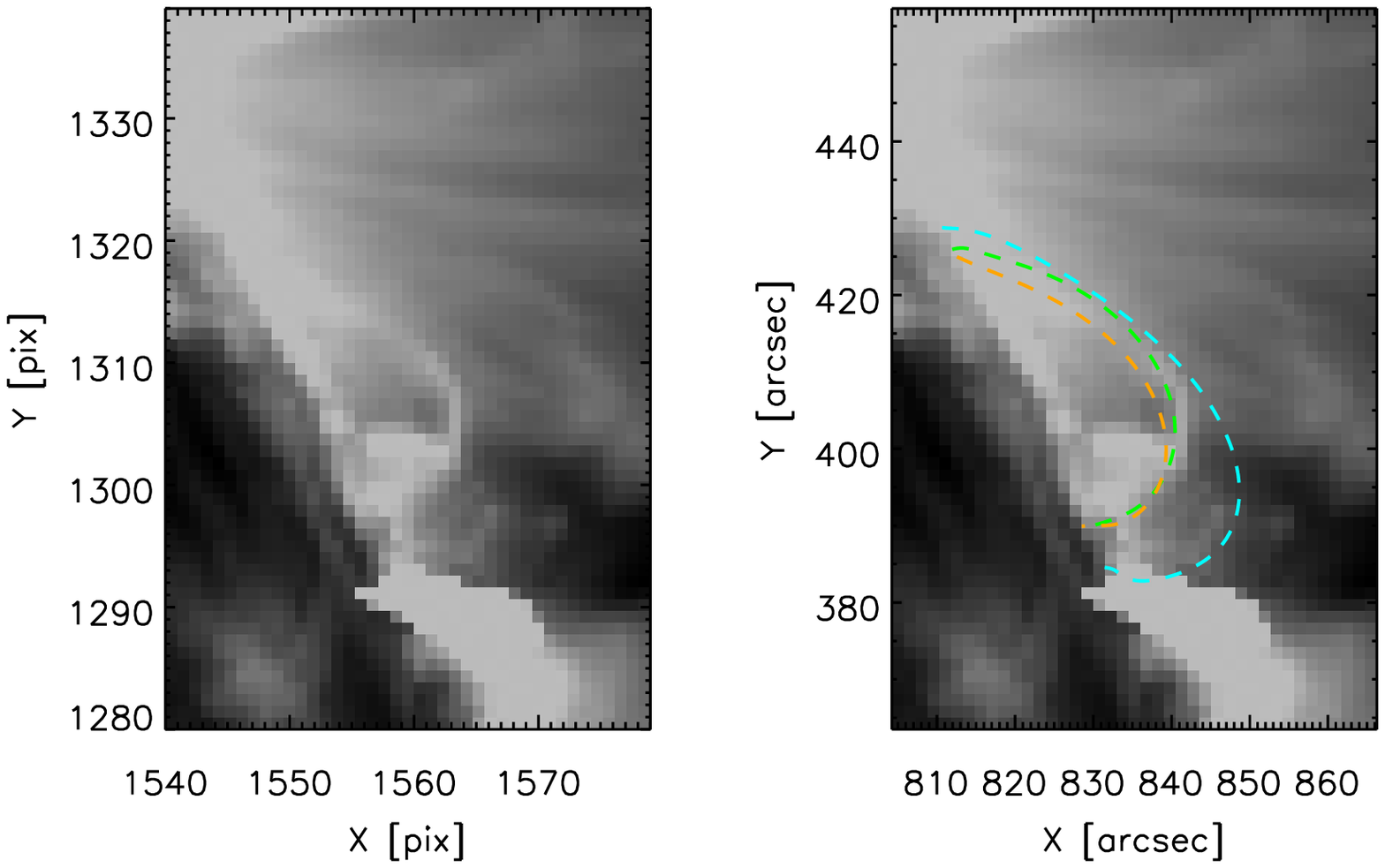}
    \caption{We call both the green and orange tracks excellent matches to the EUV feature\change{s observed with EUVI-B $171${\AA}}, despite the possibility for ambiguity between them.}
    \label{fig:example1}
  \end{figure}

\begin{figure}[htbp]
\centering
     \includegraphics[width=0.999\linewidth]{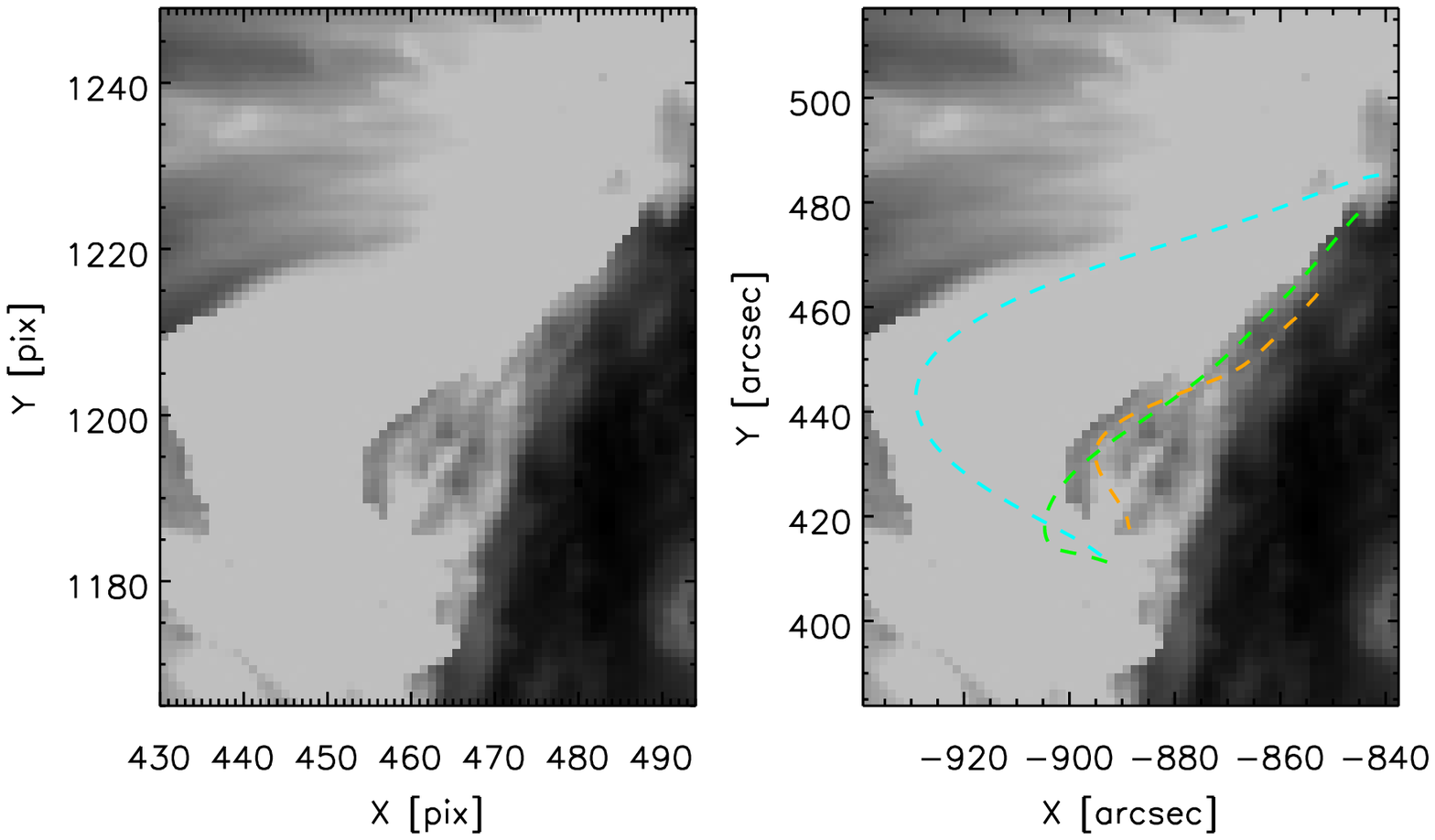}
    \caption{\change{Two} tracks in this example are labeled maybe matches \change{to features in EUVI-A $171${\AA}}.  The green track lies on the low-altitude edge of the saturation.  The orange track also is on the low-edge of the saturation, and there are some bright features to which it conceivably aligns.  \change{The cyan track lies believably within the saturation, but is removed from our statistics as features it might correspond to cannot be resolved.}}
    \label{fig:sat}
  \end{figure}

\begin{figure}[htbp]
\centering
  \includegraphics[width=0.999\linewidth]{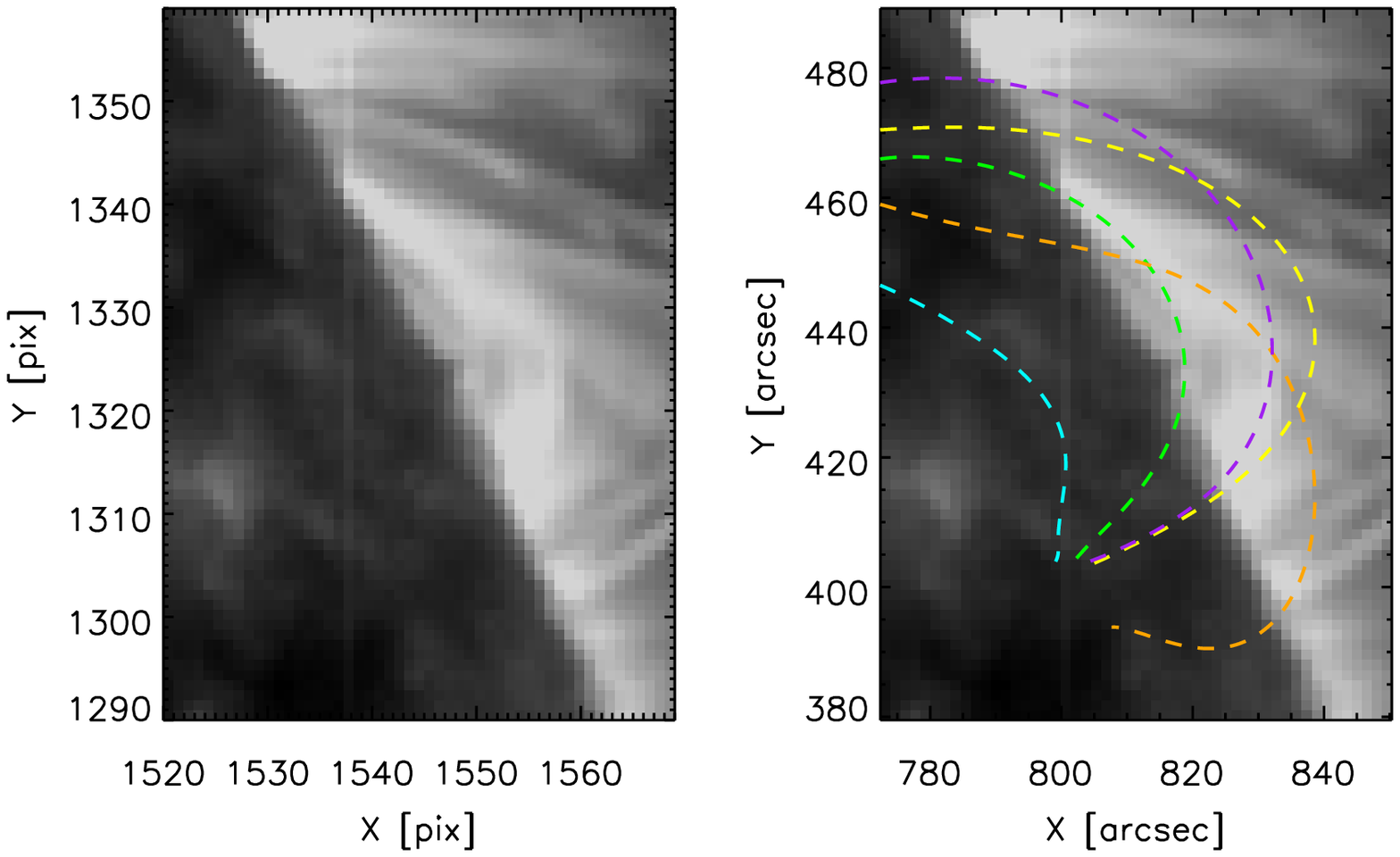}
  \caption{At this point in time, the feet of the loops lie on the other side of the limb.  Therefore, the low-altitude portions of the tracks (including the entire cyan track) are obscured due the disk.  Tracks like the cyan loop are then excluded from consideration in the visual analysis. }
    \label{fig:limb}
  \end{figure}

To assess the fidelity of the $\alpha$-h fitting, we compare the tracks with the EUVI observations to determine how well the two match in this different viewing angle.  We first perform a qualitative assessment of all loops by visual inspection.   The results are summarized in this section with some key examples for context\footnote{The examples in Figures \ref{fig:example1}, \ref{fig:sat}, \ref{fig:limb}, \ref{fig:example2}, and \ref{fig:example3} have been square root-scaled with varying values for the color scale saturation for clarity purposes only.  The subsequent data analysis in \secref{sec:quant} uses intensities in units of DN/s.}.  \change{Figures \ref{fig:Aexample} and \ref{fig:Bexample} show instances of tracks plotted on an EUVI-A and EUVI-B image, respectively}.  %and \figref{fig:Bexample} shows an instance of tracks on one from EUVI-B.}  
The full set of all EUVI images with the tracks over-plotted %\change{(18 images for EUVI-A and 26 for EUVI-B)} 
can be found in the supplemental figures \change{in the online journal}\footnote{Note that in the supplemental figures, the tracks are directly plotted over the EUVI image.  As a result, in some of these figures (earlier in time for STEREO-A and later for -B) part of the track that would be behind the limb is shown on the image and thus some observer intuition is required.}.  Some $\alpha$-h models were unable to be examined due lack of EUVI data availability around the time of interest.  (This is not explicitly shown in the subset of the data used to for illustrative purposes in \figref{fig:times}, but would be a case of a vertical dotted line that has no red or blue vertical lines near it.)  However, some models could also be viewed from both STEREO satellites.  Thus, of the 199 tracks that were constructed, there are 173 to compare to EUVI images.  \changetwo{We seek the closest underlying feature in the images to pair with each track.  \changethree{That is, we did not consider using one track to compare to more than one loop in EUVI.  As a result, each track only gets put into one goodness-of-fit category per image}.}    We define three categories in which we classify the goodness of fit: excellent match, maybe match, and no match.  Note that one track can be placed both into the no match category when viewed in EUVI-A while still being an excellent match in EUVI-B, for example; we found 33 tracks with some variation of this.  There were a total of 173 tracks that coincided with the EUVI dataset, with 8 that could not be compared to the data (i.e., some loops were entirely hidden behind the limb).  \change{As some saturation in EUVI data made fine structure unresolvable, 14 tracks were further removed from consideration due to tracks being entirely within the saturated area.}  Of the \change{151} tracks that were viable for visual comparison, \change{40 ($26.5\%$)} were excellent match, \change{65 ($43\%$)} were maybe match, and \change{46 ($30.5\%$)} were no match.

%\changetwo{\colorbox{cyan}{Here we talk about the 1:1 correspondence between a track and a loop}}

\change{\input{figset}}

To assign a track to one of our goodness-of-fit categories, we used the following reasoning.  For an excellent match, a track would follow a loop in EUV along the loop's axis over at least enough of the visible structure for it to be convincingly and appropriately compared.  In \figref{fig:example1}, we classify both the green and orange tracks excellent match from their close alignment with a coronal loop structure.  This is despite the ambiguity in resolving the two structures individually.  
Since the two tracks are so close in this perspective it is possible that the visible loop includes both, or only one.  

If the track was on the low- or high-altitude edge of the structure, or otherwise did not align completely with the observed loop, we called this a maybe match.  Maybe matches were also assigned if the track fit within saturation on the image reasonably but could not be matched to finer structure.  For an image with saturation in the region of interest, like in \figref{fig:sat}, we assign the three tracks into the \change{following categories}.  %maybe match category.  
The cyan track fits believably within the saturation but is otherwise unable to be further resolved \change{and was removed from consideration}, while the others (green, orange) \change{are considered maybe matches as they} trace the low altitude edge of the saturation and correspond to some of the other bright features.  \change{Other maybe match examples are also shown in the yellow tracks of Figures~\ref{fig:Aexample} and \ref{fig:Bexample}.  These tracks are not quite aligned with the EUV structure, but are aligned enough to be maybe a match.}

The no match category was populated with tracks for which we did not see EUV features to which they corresponded.  Not included in any category, nor in the total count, were tracks whose position relative to the line of sight made it impossible to compare them with structure in the corona.  An example can be seen in \figref{fig:limb}, where the low-lying cyan track lives on the far side of the limb entirely and the disk blocks any view of it from this LOS.  As in any sort of qualitative analysis, there is a level of subjectivity involved.  Thus, we invite the reader to look at the supplemental figures and draw one's own opinion.

In general, the co-alignment of the tracks to the EUVI observations fit well.  Together the set of excellent matches and maybe matches exceed the amount of no matches.  One might expect more discrepancy between the model and observation due to the lack of temporal precision between the data from AIA (used in the fitting) and from EUVI (used to observe); that does not seem to be the case.  We also find that the EUVI data (particularly from the A spacecraft) are very saturated.  Without saturation in the data set, we might be able to turn maybe matches into excellent matches.  

 We find that certain types of loops are particularly well fit by the $\alpha$-h fitting method.  These loops share a similar-looking morphology, appear to have their feet anchored at similar locations and rise to similar heights in the atmosphere.  The green and orange tracks in \figref{fig:example1}, the purple in \figref{fig:example2}, and orange in \figref{fig:example3} are examples of this well-fit group.  Their particularly good matching might be explained by the general topological structure of the 3D field.  The aforementioned loops all lie in a similar location and, more significantly, they belong to the same magnetic domain identified in the magnetic modeling of \citet[][see Figure 4 of that work]{Longcope2020}.  These particular tracks connect between the polarities identified by \citet{Longcope2020} as P06, 07, 08 (adjacent parts of the same photospheric flux) to N10 (see the purple track in \figref{AIAex}).  The flux partitioning may occur in such a way that the domain which contains these morphologically similar loops has a very good fit to the LFFFs. 

\change{A loop's inclination with respect to the local vertical allows for the possibility that there is some dependence on viewing angle on how well it is observed.  Following this line of thinking, there might be some trends between inclination angle of the track and the goodness-of-fit category in which it was placed.  We tested this by calculating the tracks' angle of inclination.  Our analysis finds no such consistency\changetwo{, as most of the tracks -- and, by inference, the loops to which they correspond -- in this study were quasi-vertical}.  The aforementioned excellent match, morphologically similar tracks had a range of approximately $5^\circ$ to $30^\circ$ with respect to the local vertical.  We found similar ranges in the maybe and no match categories as well.}%There exists the possiblity...  However we find t}
%\change{\colorbox{cyan}{Some analysis of inclinations?} Well, they're all similarly in the vertical...  Whether that speaks to the fitting algorithm or something else hmmm well! inclination of loops might explain, if we were to expect this.  Analysis of the incliatino of our tracks reveal no such consistency.  Both excellent fits and no matches can have $\sim$6 degree angle wrt verticl, or larger 30-40 degrres.  No consistency, wide range}

\section{Quantitative analysis of loops using Gaussian fitting}\label{sec:quant}

Following the qualitative assessment of alignment we select a subset of high-quality examples for more quantitative analysis in regards to the loop location and width.  We would like to compare these quantities more precisely.  We develop and apply a methodology in order to overcome analytical difficulties within our data.  First, though we know the track's location in the POS, we do not know the exact axis of the loop and therefore its location in the POS as defined by the EUV emission.  Additionally, we would like to measure both the loop's location and its width without bias or error from an incorrect guess as to the axis direction as informed by the track.  \change{For example, it is likely the direction perpendicular to the track is not the width we would like to measure for a misaligned track/loop in the maybe match category.}

\begin{figure*}[htbp]
\centering
     \includegraphics[width=0.72\linewidth]{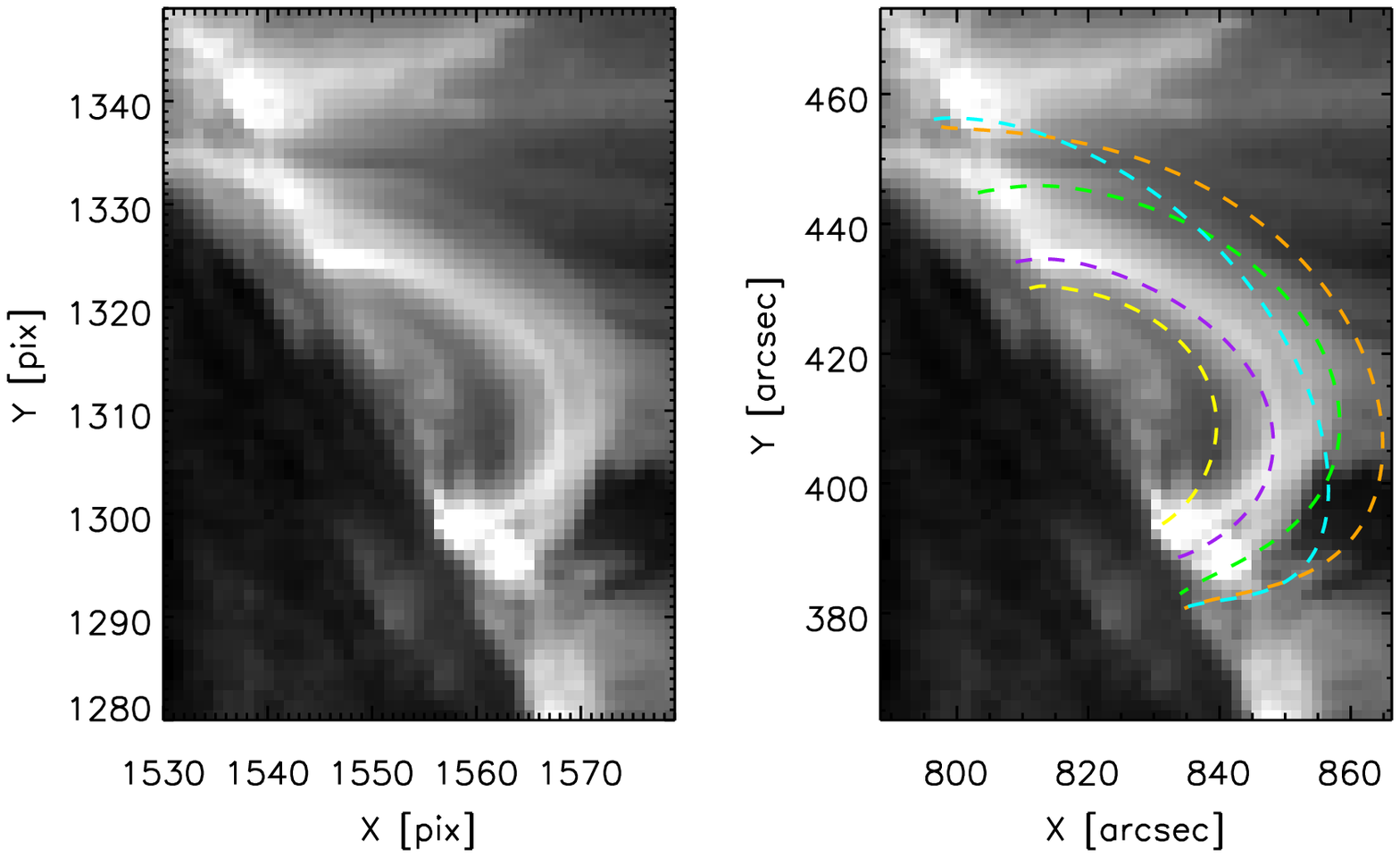}
    \caption{This instance of $\alpha$-h fitting resulting in these tracks is denoted by the purple star in \figref{fig:times}.  The purple and green tracks were categorized as excellent matches and trace out loop structures, though the corresponding feature to the green track is fainter than that which matches the purple.  The purple trace and the EUV feature to which it corresponds is referred to as Example 1.  Note that the colors in this figure correspond to the same track colors in \figref{AIAex}.}
    \label{fig:example2}
  \end{figure*}

\begin{figure*}[htbp]
\centering
     \includegraphics[width=0.72\linewidth]{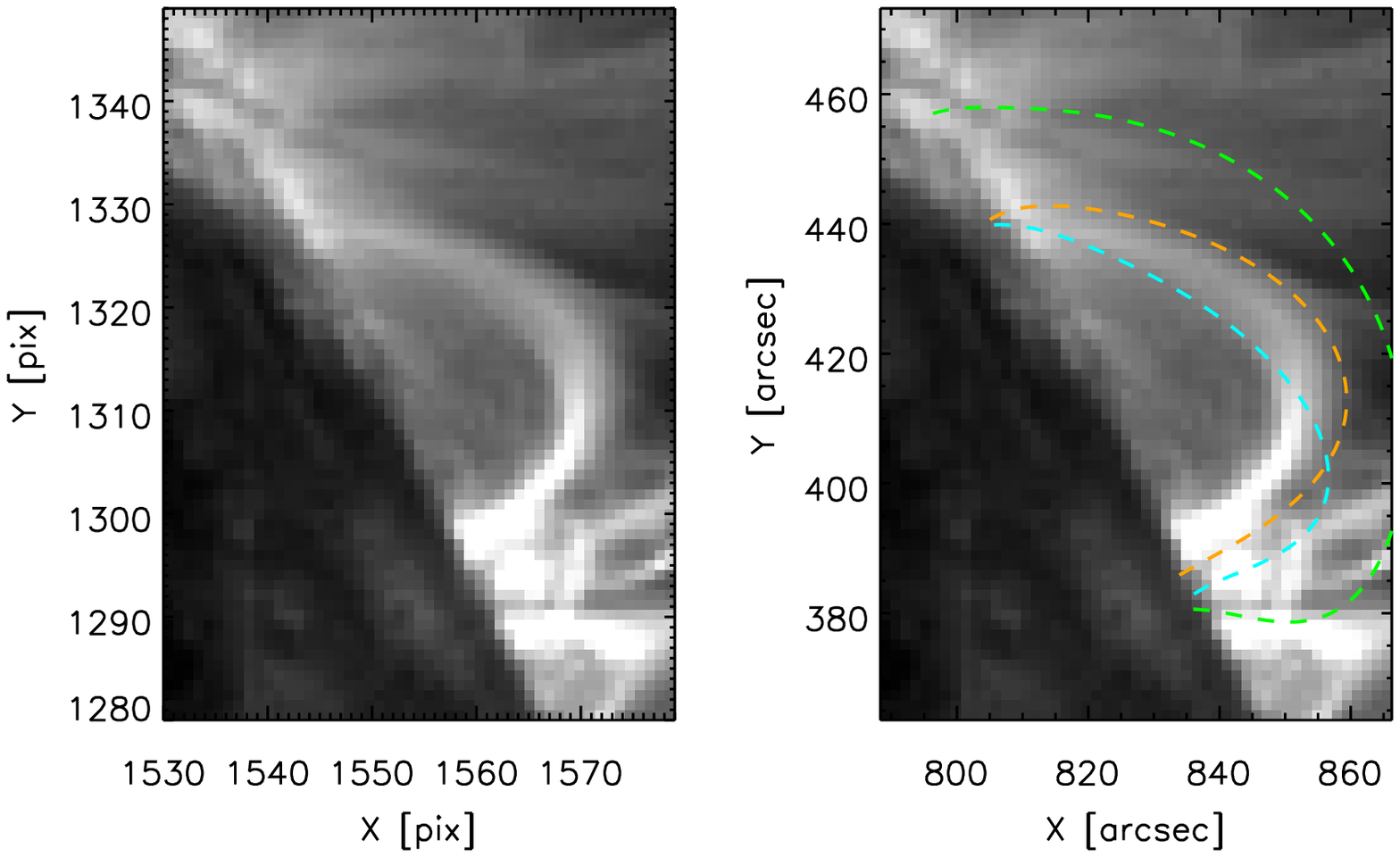}
    \caption{Both the cyan and green tracks this example belong to the no match category.  The orange track is determined to be a maybe match.  Though it does not align completely along the loop's axis, it consistently traces out the high-altitude edge of the feature.  We refer to this orange track as Example 2.  In \figref{fig:times}, this time is marked by the orange star.}
    \label{fig:example3}
  \end{figure*}

Two particularly illustrative examples are given in Figures \ref{fig:example2} and \ref{fig:example3}.  The example in \figref{fig:example2} assign the purple and green tracks to the excellent match category, orange is a maybe match, with the yellow and cyan assigned to no match.  The purple track (hereafter called Example 1) and its underlying feature are particularly well suited to detailed quantitative analysis.  \figref{fig:example3} has two no match (cyan and green), and one maybe match (orange).  The orange track seems to lie on the high-altitude edge of a coronal loop, and we will investigate this in further detail as well.  We call the orange track Example 2.  In \figref{fig:times}, the instance of Example 1 is denoted by the purple star and the orange star is that of Example 2.

The data set contained a small sample of 11 tracks that we considered suitable for comparison to the observed loops in this sort of quantitative analysis.  This was due to issues with the images in EUVI, including saturation and overlapping bright features making a particular trace not resolvable.  Of these 11, 9 were observed with from one STEREO and AIA and 2 were observed from both EUVI and AIA simultaneously.  In total, 13 track/loop pairs were examined.  \changetwo{In this subset of the data, there is a one-to-one correspondence between a track and the matching loop. }%As before, were there was at most one loop observed in EUV that we consider corresponding to a track.  That is, we seek the closest underlying feature in the images to pair with each track.  There is, however, occasions where there is no observed matching feature.

\subsection{Methodology}

\begin{figure*}[tbhp]
\centering
	\includegraphics[width=0.999\linewidth]{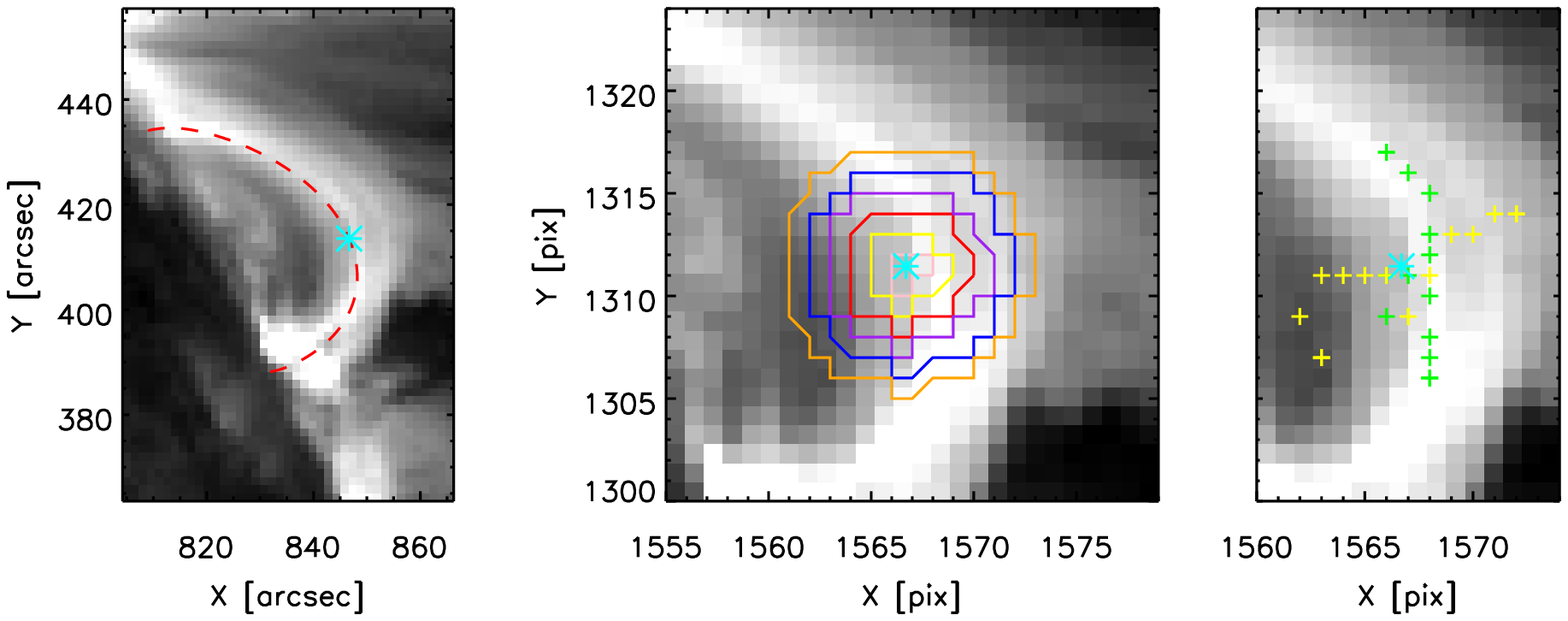}
                  \caption{Left: The track (red line) shown here is the same as Example 1 (\figref{fig:example2}, purple).  A particular point of interest along the track is denoted by the cyan star. \\Center:  Six concentric \change{shapes} %circles 
                  are shown.  \change{These shapes are determined from drawing a circle, then matching the perimeter of the circle to the nearest pixel.  These pixels}  %The pixels along the perimeter of these \change{shapes} %circles 
                  are the ones whose intensities are used to find the maxima and subsequent edges\\Right:  The maxima along the perimeter of each of the six circles are shown in green crosses and the corresponding edges are plotted in yellow crosses. }
    \label{fig:step1}
  \end{figure*}

To address the concerns described above, the quantitative fitting for each EUV loop (which correspond to a track) is conducted as follows.  We begin with a point in the track called the point of interest (POI, see the cyan star in \figref{fig:step1}).  We then identify a circle of radius $r$ and extract the intensity values \change{from the pixels} along its perimeter.  From this array of values we identify two local maxima as the axis or spine of the loop.  Between pairs of maxima, on both sides, we then find the minimum intensity value in this range.  These two minima we call the ``edges.''  

We repeat this for circles with radii in the range 1 to $n$ pixels.  The number of circles used, $n$, varied for each track based on the data around it.  In \figref{fig:step1}, the middle panel shows six of these circles over an EUVI image with the cyan star designating the POI.  The right panel of the figure shows the POI with the maxima locations plotted in green crosses and the edges plotted in yellow crosses.

\begin{figure}[htbp]
\centering
     \includegraphics[width=0.999\linewidth]{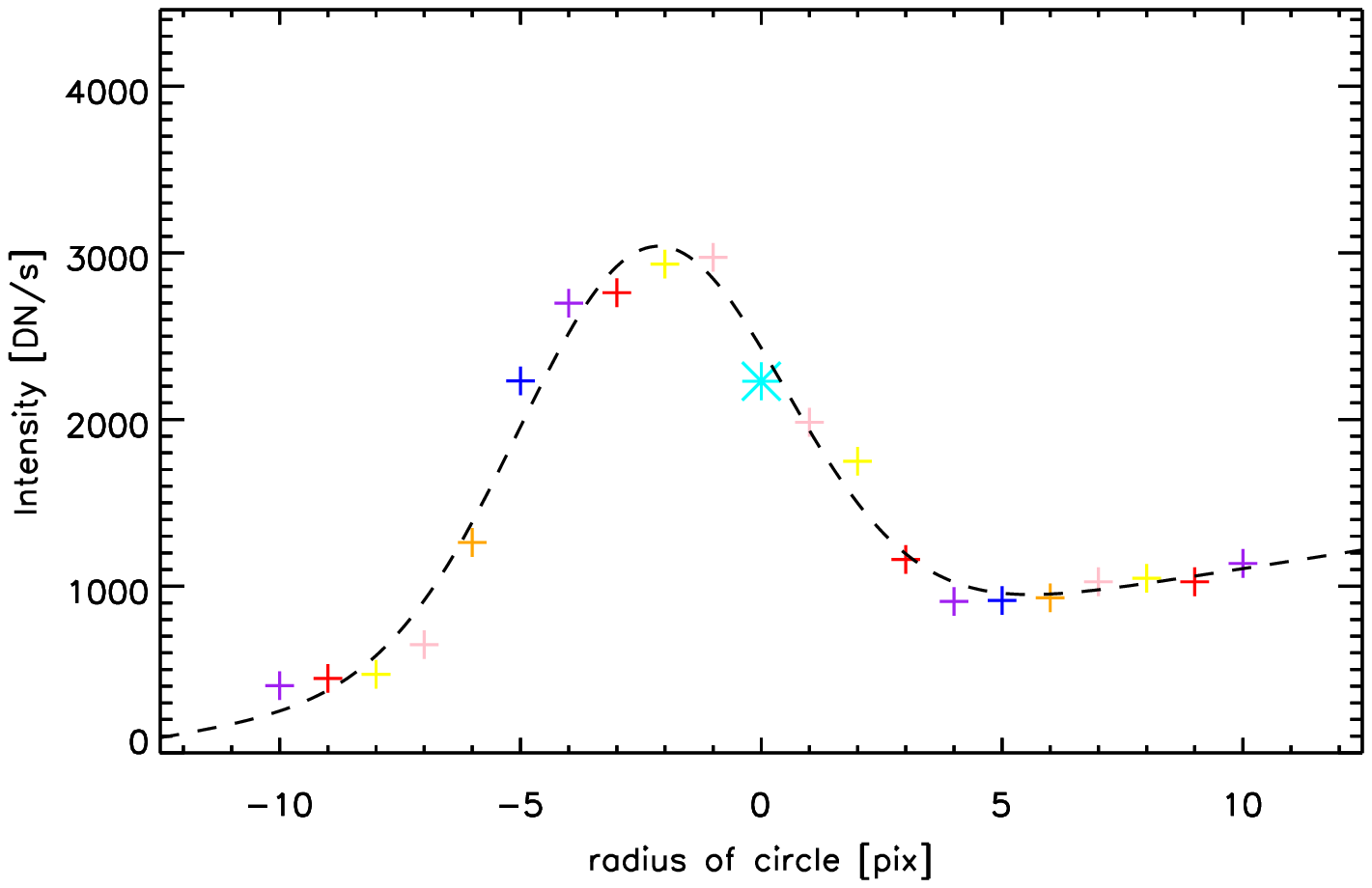}
        \caption{An example plot showing one instance of the intensity of the edges (crosses) and the POI (cyan star) vs. radius of circle used (different colors).  This particular instance used 10 concentric circles and their edges.  These data were then to fit a Gaussian (dashed line). }
    \label{fig:step3}
  \end{figure}

The intensity values of the edges are then used to perform a Gaussian fit.  As in \figref{fig:step3}, for all the edges we plot intensity value of the edges vs radius of the circle used.  The intensity of the POI is also on this plot at $r=0$, plotted again in a cyan star.  Each circle has two edges which we designate positive and negative, and hereafter assign position $x=+r$ and $x=-r$.  We perform the Gaussian fit to this plot using the IDL function {\tt GAUSSFIT}.  Five terms $(A, B, C, D, E)$ were used for the Gaussian such that
\begin{equation}
f(x) = A e^{-\frac{\left(x-B\right)^2}{2C^2}}+D+E x
\end{equation}
and with initial estimates of $A=$ intensity of POI, $B=0$, $C=2$, $D=$ mean of all the intensities, $E=0$.  The estimates aided {\tt GAUSSFIT} in fitting the Gaussian near the point of interest as opposed to centering the fit at larger values of $r$ that may have large intensity values.  The large intensities at larger radii are likely to be other structures, as we expected the loop to be centered around the point of interest.  The centroid is given by the parameter B, which often differs from B=0 (the original POI).  The full width at half maximum (FWHM) of the fit is what we call the loop diameter and is computed $$2\sqrt{2 \ln\left(2\right)}C.$$  

This process -- which has been described for a single POI -- is repeated along the length of the track.  As each track is defined by a set of $(x,y,z)$ coordinates, each point along the track has its corresponding location in both the AIA and EUVI FOVs.  We progress through all points that define the track, doing the Gaussian fitting using that point as the POI.  For each point along the track, we have multiple values for observed diameter.  Each value is from a different spacecraft/LOS, but they all correspond to the same 3D location along the track.

\begin{figure}[htbp]
\centering
     \includegraphics[width=0.999\linewidth]{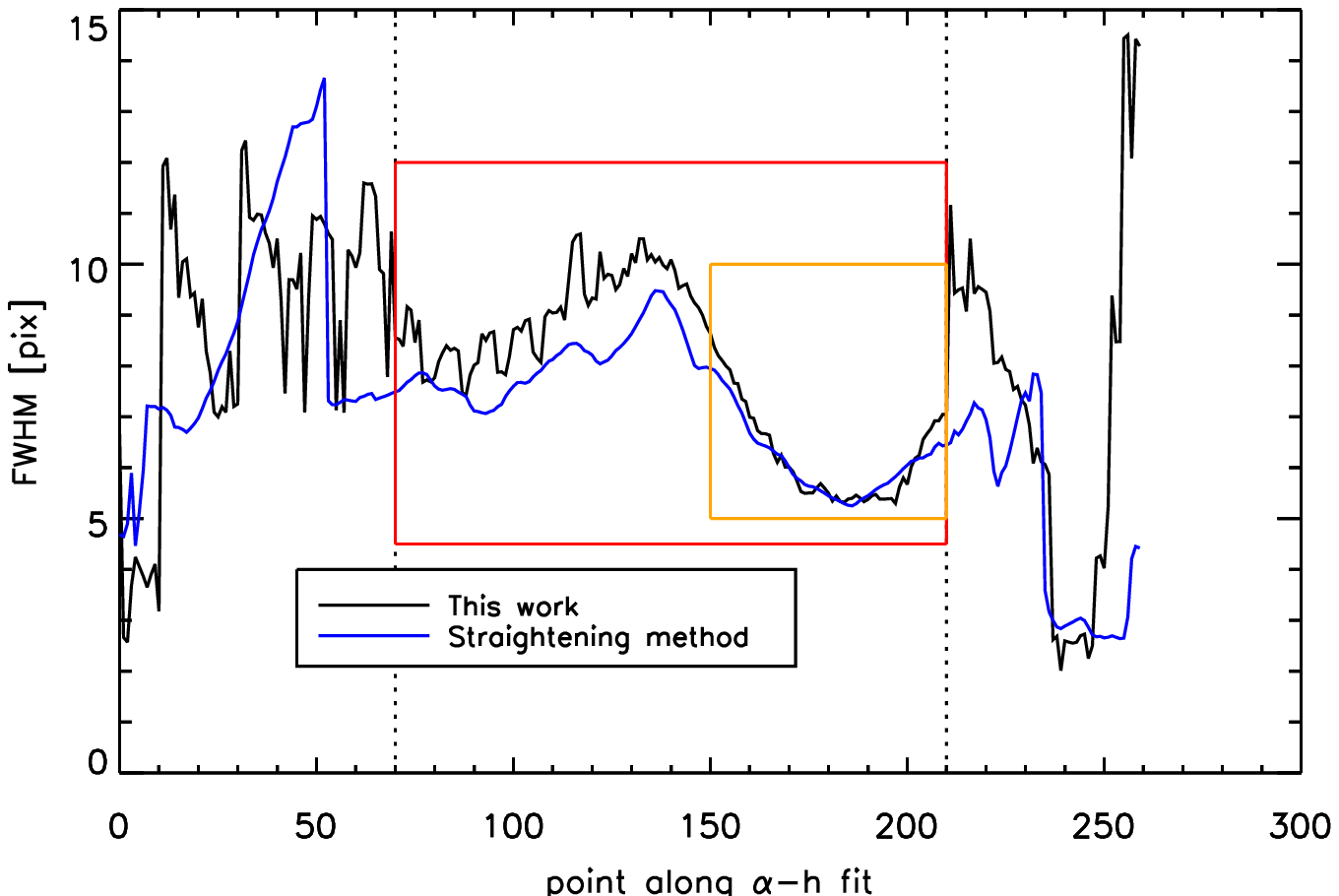}
     \includegraphics[width=0.999\linewidth]{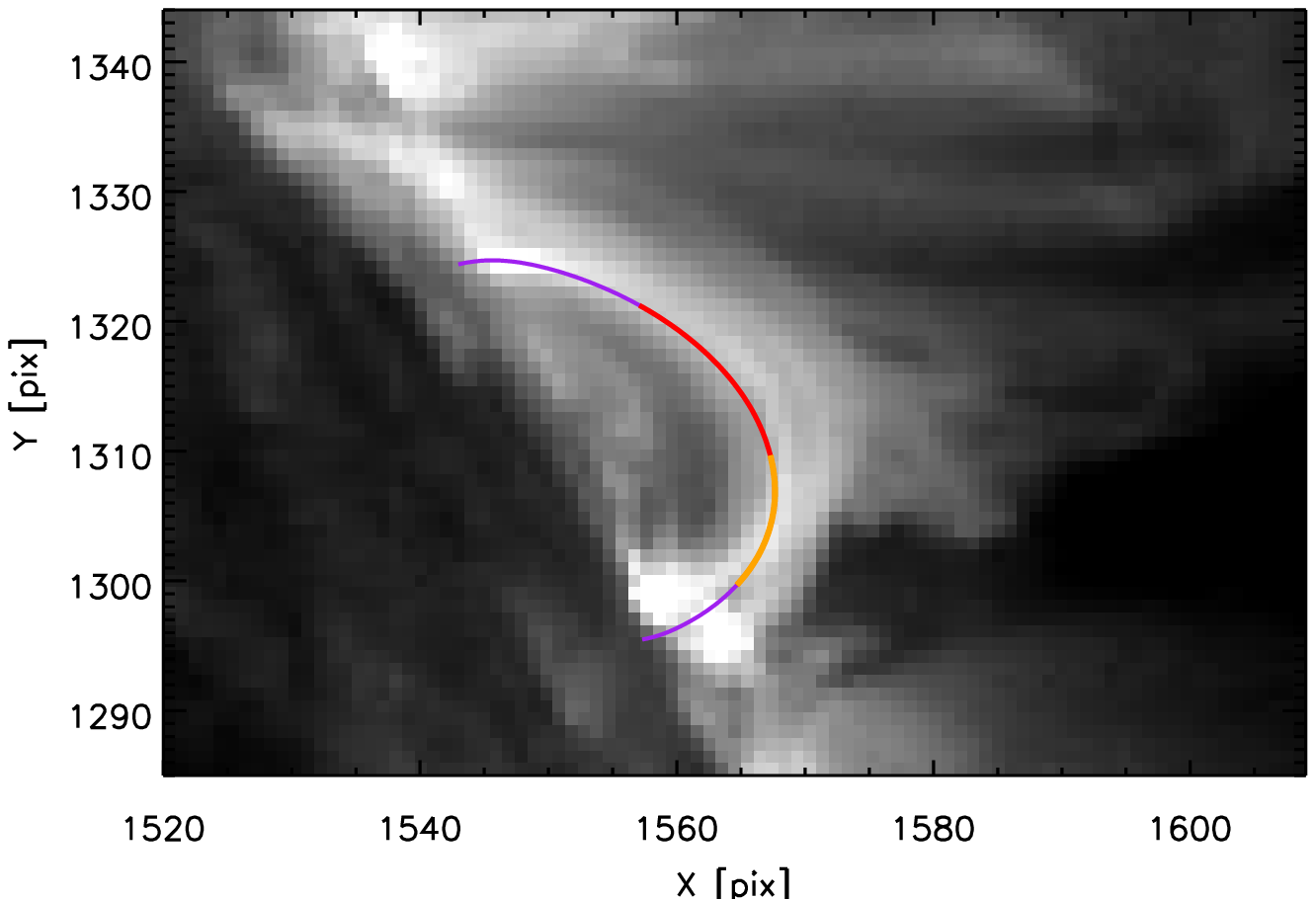}
        \caption{\change{A comparison of this work's concentric-circle method (black) with that of straightening the loop (blue).  The red-boxed region is the original region of good fit, and the orange box encloses a region of excellent agreement between the two methods.  The regions in the red and orange boxes in the top panel correspond to the same color of the track in the bottom panel.}}
    \label{fig:straightcompare}
  \end{figure}

\change{We find that this method is comparable to that of ``straightening'' the loop.  In such analysis \citep[as in][]{2000SoPh..193...77W,2003ApJ...593.1164W,2008ApJ...680.1477A}  a curve is traced on the image and then the data is interpolated into an array in which the coordinate system has one vector that is locally parallel to the curve and the second is perpendicular to the first.  \figref{fig:straightcompare} shows the diameter comparison between the two methods using similar parameters.  The track was used as the curve to which the coordinate system ran parallel and perpendicular.  The same Gaussian fitting method was performed for each row (intensity profile) in the interpolated array as in our concentric-circle method.  There is good agreement between the two in the well-fit region.  (The determination of this well-fit region is detailed in the next section.)}  

\changethree{Our method provides us with two advantages over this other method.} First, we do not have to interpolate the imaging data. \changethree{Second, our method allows us to use our theoretically derived tracks, without requiring the identification of a more accurate loop axis independently.  A track is based on a magnetic model of the field line, and found in \secref{sec:byeye} to match reality imperfectly.  Therefore, in order to apply the other method, we would need to take the additional step of tracing the actual loop axis through each image, defining a curve along the loop to be straightened.}%   We base our method on a set of tracks obtained from a magnetic model of the field line which is found in \secref{sec:byeye} to match reality imperfectly.}

\begin{figure*}[htbp]
\centering
     \includegraphics[width=0.49\linewidth]{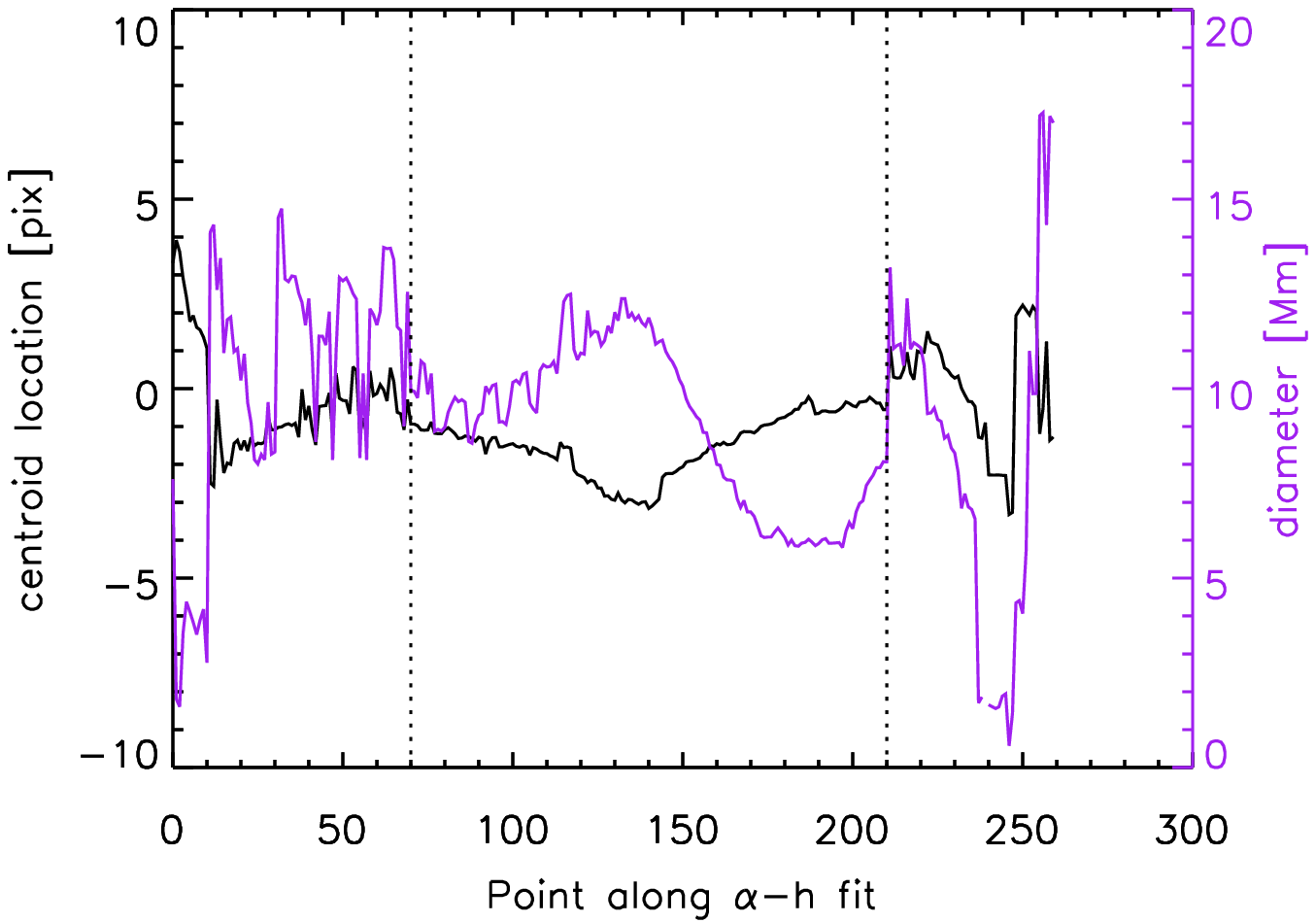}
          \includegraphics[width=0.49\linewidth]{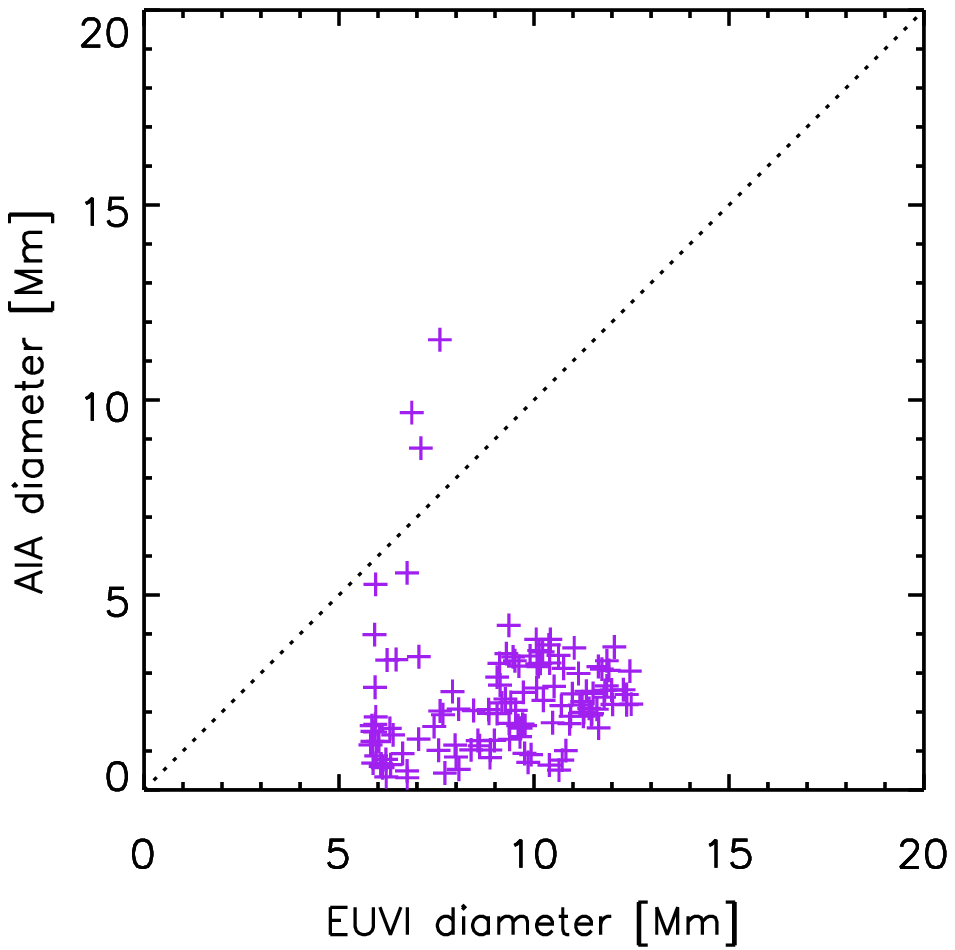}
        \caption{These plots correspond to the quantitative fitting methodology applied to Example 1 (purple loop in \figref{fig:example2}). Left: A plot from EUVI data of centroid deviation from POI location (black) and diameter of the loop (purple) as a function of point along the track.  The region between the dotted lines is the best fit region.  Right:  Scatter plot of EUVI diameter vs AIA diameter for the points within the best fit region.  The statistic Spearman's $\rho = $\change{0.226} $\mathchange{(p = 7.1\times10^{-3})}$ for this distribution.}%\change{\colorbox{cyan}{p value:} 0.00712}}
        \label{mod10fits}
\end{figure*}

\begin{figure*}[htbp]
\centering
     \includegraphics[width=0.49\linewidth]{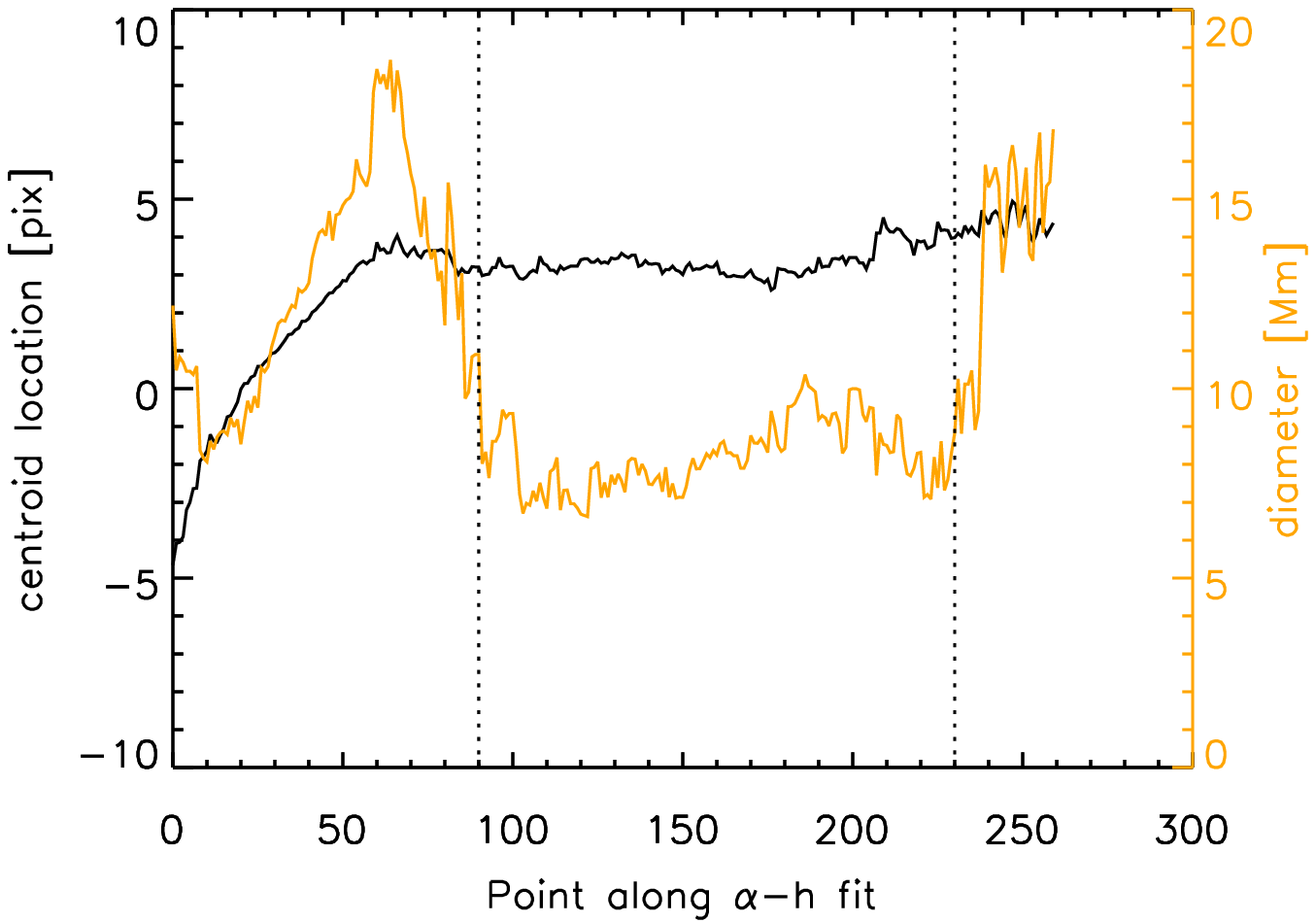}
          \includegraphics[width=0.49\linewidth]{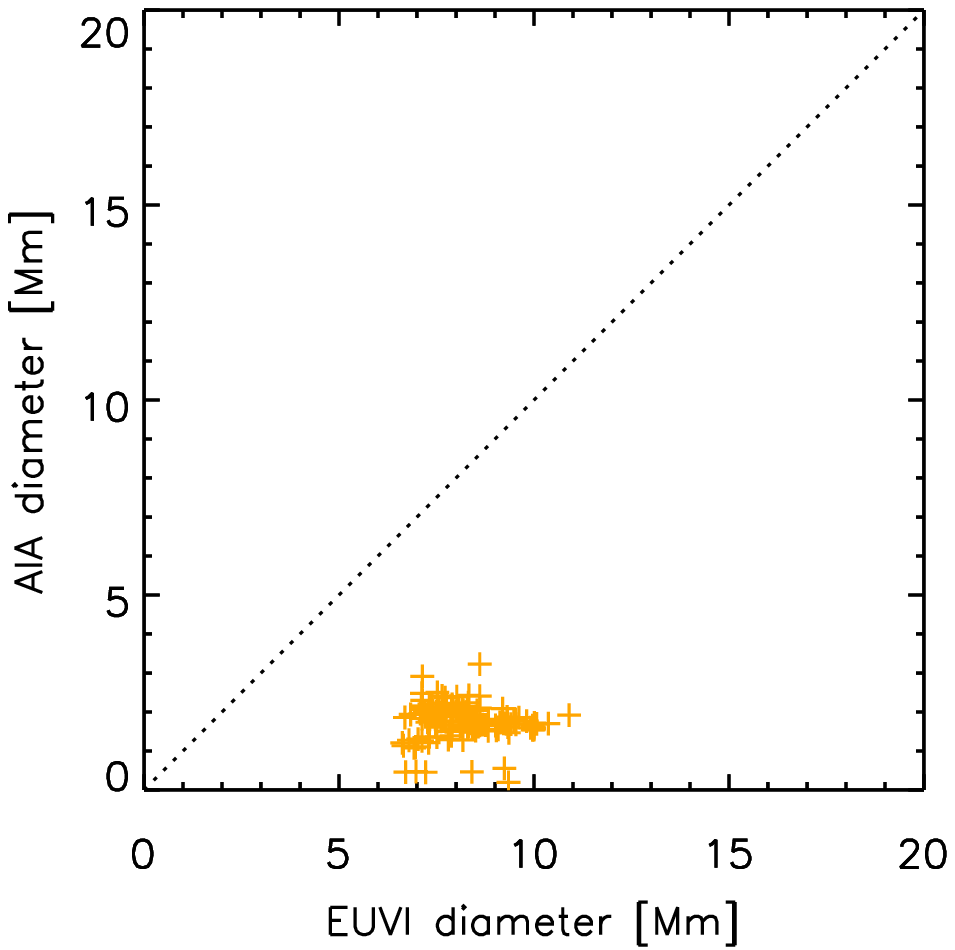}
        \caption{These plots correspond to the fitting method applied to Example 2 (orange loop in \figref{fig:example3}). Left: Centroid deviation from POI (black) and FWHM loop diameter (orange) as a function of point along the track as viewed from EUVI.  The region between the dotted lines is the best fit region.  Right:  Scatter plot of EUVI diameter vs AIA diameter for the points in the best fit region.  The Spearman's $\rho $ statistic is $-0.13$ $\mathchange{(p = 0.1)}$ for this distribution. }
        \label{mod13fits}
\end{figure*}

\subsection{Centroid}

We use the centroid of the Gaussian fit to assess how well the tracks fit to features in EUVI.  This analysis bolsters and quantifies the original visual evaluation.  The position $r=0$ (see \figref{fig:step3}) corresponds to the POI, located where the AIA track is projected.  If the Gaussian fit returns parameter $B=0$, then the axis of the EUVI loop coincides with the POI exactly.  Any non-zero value indicates a discrepancy.  The centroid value was also used as a proxy in determining how well the Gaussian fit the data.  For example, significant fluctuations in the centroid value (as a function of location along the track) indicated that the Gaussian fitting was not locked onto a consistent structure.  We see this sort of fluctuation for Example 1 in the left panel of \figref{mod10fits}, at both ends of the track.  The companion image (\figref{fig:example2}) shows bright features, which are not necessarily the loop in which we are interested, around the feet of the track.  In the region the track was well fit (denoted by the region between the vertical, dotted lines in \figref{mod10fits}, left), the median deviation from the point of interest was approx. $ -1.1$ EUVI pixels%\colorbox{cyan}{(about $1.75$ arcsec)}
.  For Example 2 (\figref{fig:example3}), the median deviation from the POI (in the region we believed the track well-fit, between the dotted lines) was about $3.2$ EUVI pixels (%\colorbox{cyan}{$5.0$ arcsec,} 
see \figref{mod13fits}, left).  For the other tracks, in the regions of believable fit, the \change{absolute} median deviation from the point of interest was in the range %$[-2.5, 4.5]$ EUVI pixels.  The median of all medians was $-0.28$ pixels.  For unsigned deviation, the range was 
$[0.04, 4.5]$ EUVI pixels\change{.  The} median of \change{all} medians \change{was} $0.77$ pixels.  \change{These values are within typical loop diameters, which is further evidence in favor of the validity of the fitting.}  This confirms that this subset of tracks were very will fit to loops in the EUVI image.  %\change{Further, these median values of derived centroids fit within typical loop diameters.  This is further evidence for the validity of the $\alpha$-h fits.}

%\change{\colorbox{cyan}{From referee's report:}  The median values of the derived centroids fit well within typical loop diameters, this is another arg in favor of fitting validity which could be mentioned **here**}

%median deviation [0.047, -0.769, -0.094, -0.364, -1.421, -2.252, 4.526, 3.234, 0.186, -0.279, 0.175, -0.819, -2.553]
%med() = -0.279
%med(abs()) = 0.769

\subsection{FWHM}\label{sec:diam}
%SDO PSF
%https://hesperia.gsfc.nasa.gov/ssw/sdo/aia/idl/psf/DOC/psfreport.pdf
%\change{\colorbox{cyan}{PSF citations?}
%sigma of 171A PSF is 1.019 pix $\approx$ FWHM 2.4 pix \citep{AIA171PSF}
%FWHM of EUVI PSF is 2.2 pix \citep{2008ApJ...680.1477A}}

\begin{figure*}[htbp]
\centering
     \includegraphics[width=0.49\linewidth]{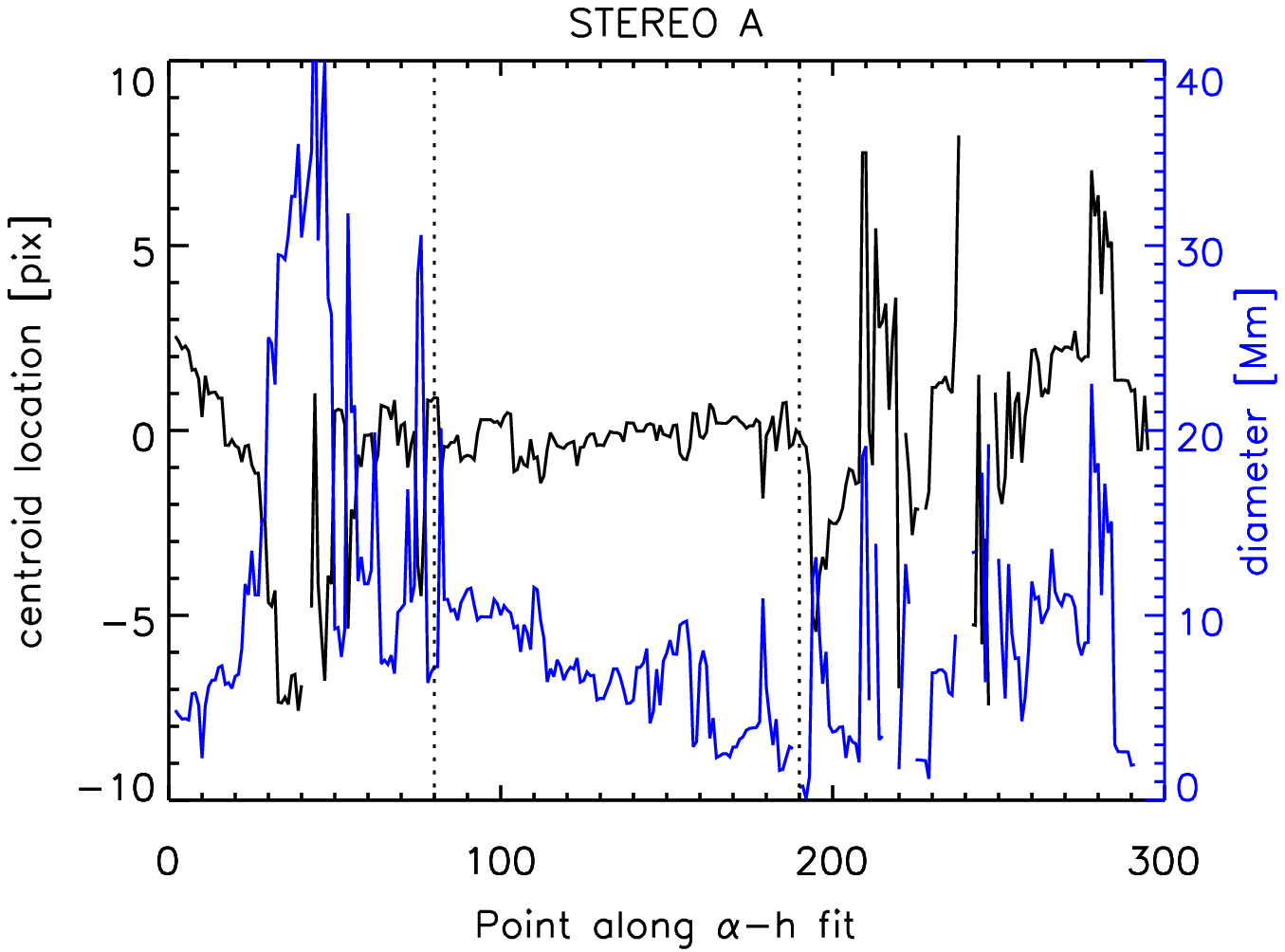}
     \includegraphics[width=0.49\linewidth]{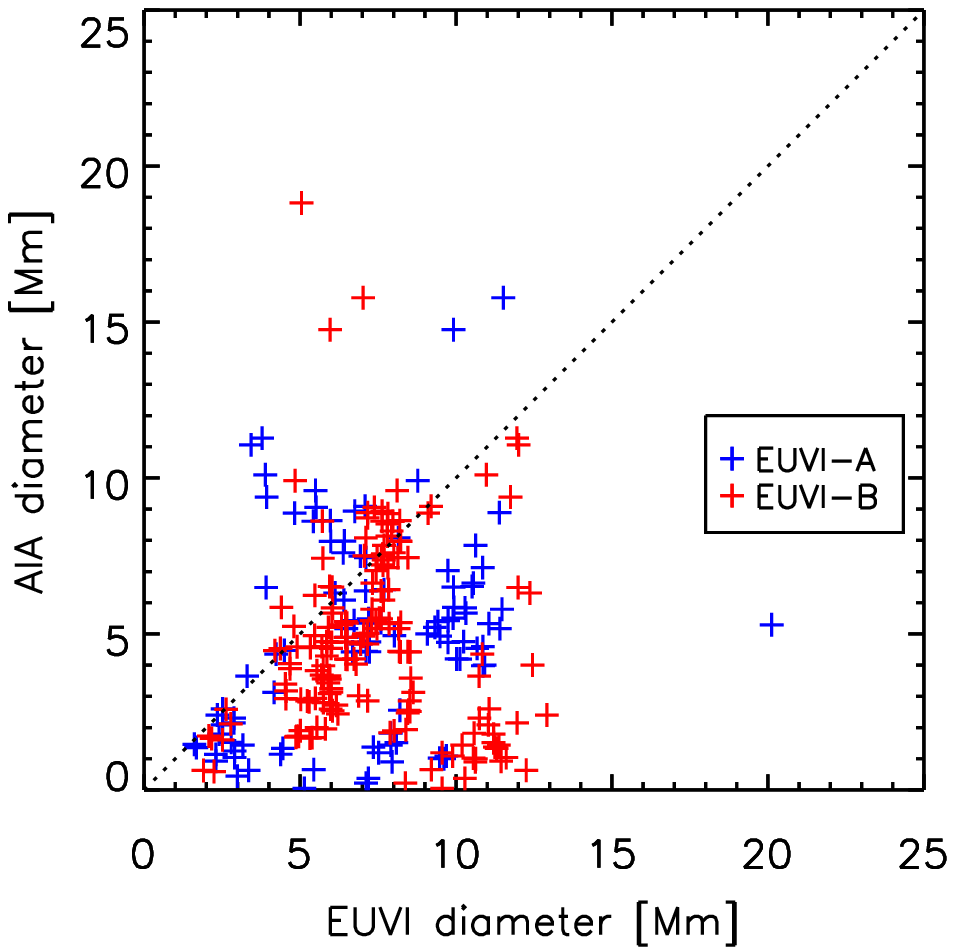}
     \includegraphics[width=0.49\linewidth]{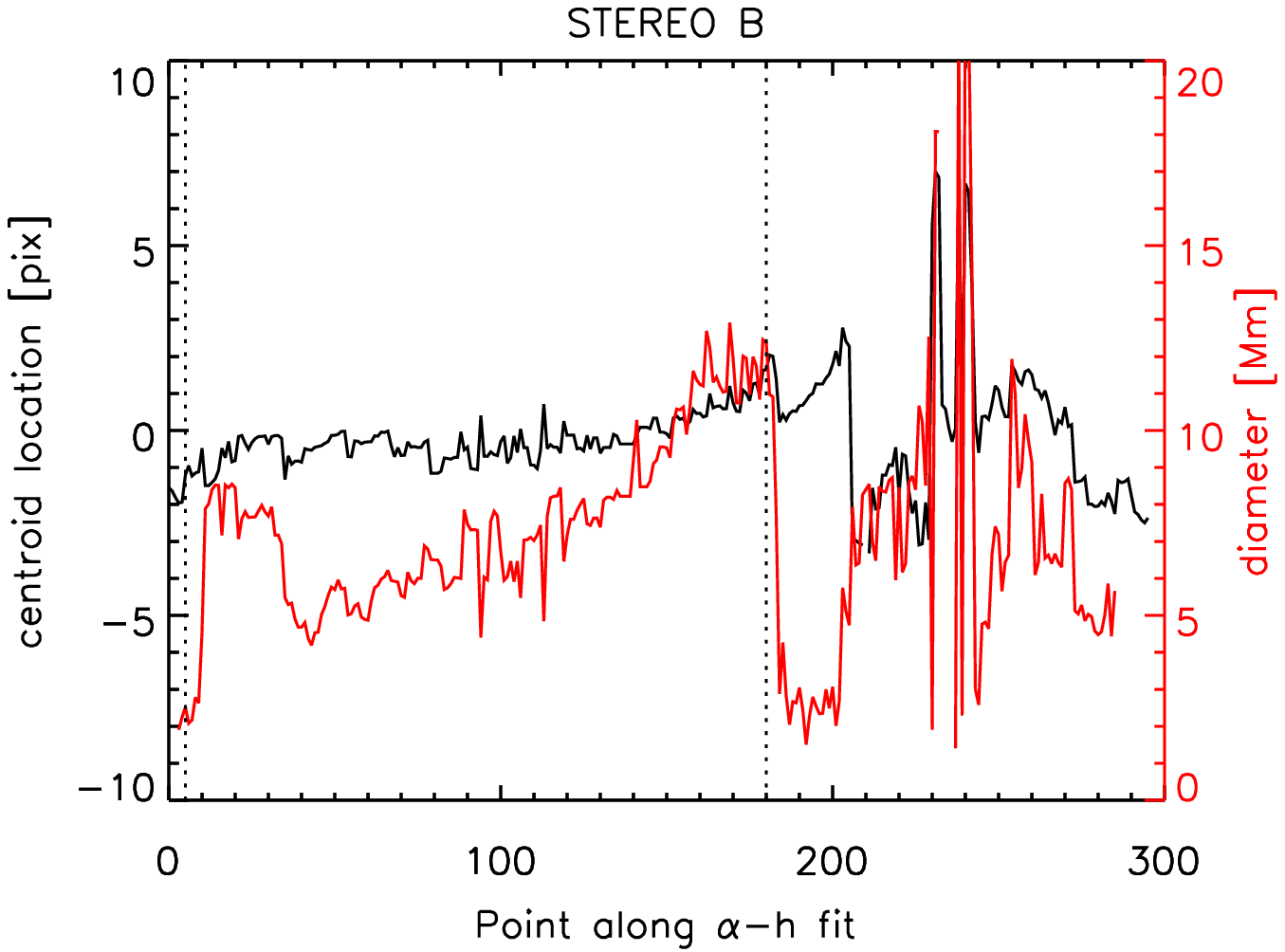}
     \includegraphics[width=0.49\linewidth]{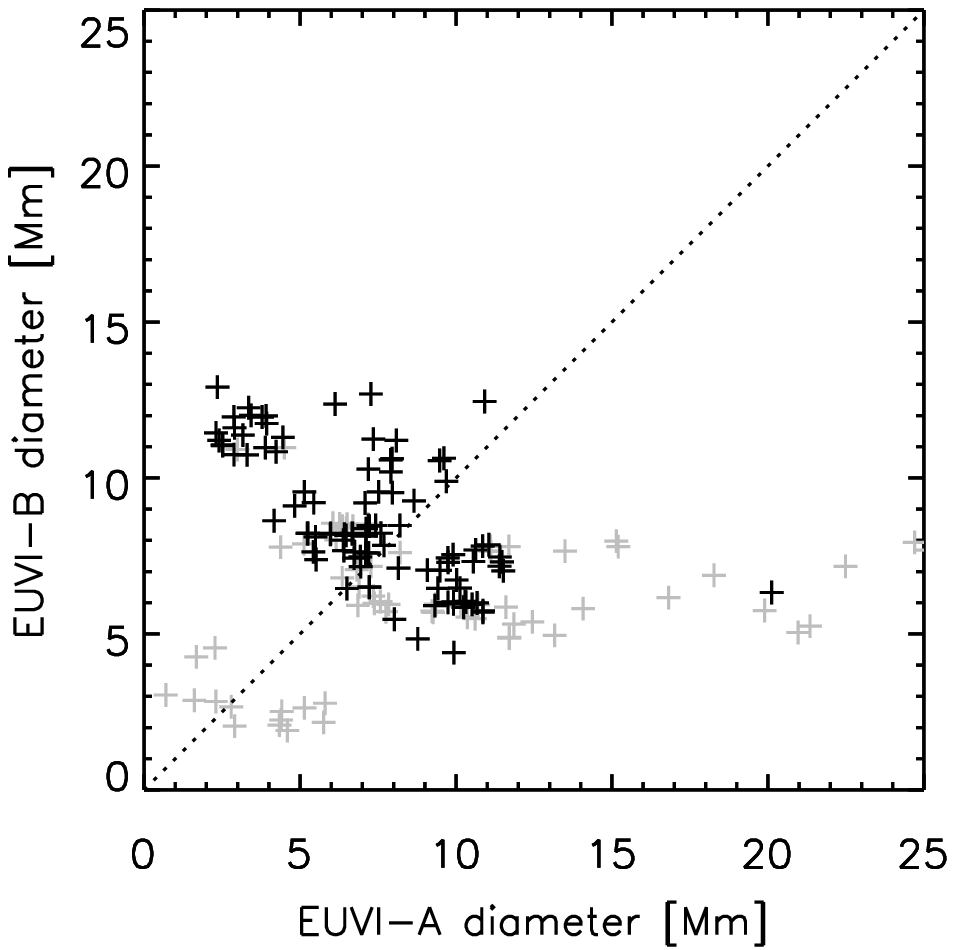}
        \caption{An example of our analysis for a loop viewed with EUVI-A, -B and AIA simultaneously.  As in previous figures, the regions are good fit are between the dotted vertical lines.  This example is \change{the yellow track shown in Figures~\ref{fig:Aexample} and \ref{fig:Bexample}} \\%{\tt model05} in the supplemental figures.\\ 
        Left: Centroid deviation (black) and FWHM loop diameter (color) as a function of distance along the track for EUVI-A (top/blue) and -B (bottom/red). \\ Top right: Scatter plot of EUVI and AIA diameters in the region between the dotted vertical lines for both EUVI-A (blue) and -B (red). \\ Bottom right: A scatter plot comparing the diameters of EUVI-A vs EUVI-B.  They grey crosses are in the well fit region in 1 instrument.  Black crosses are POIs that are well fit in both.
        The statistic Spearman's $\mathchange{\rho = -0.392}$ $\mathchange{(p = 2.05\times10^{-8})}$ \change{for all crosses, and} $\mathchange{\rho = -0.647}$ $\mathchange{(p = 2.55\times10^{-13})}$ \change{for black crosses only}}%-0.406$ (all crosses), $-0.647$ (black only).}
\label{fig:dual}
\end{figure*}

%fwhm ratios [1.717, 1.902, 1.582, 1.661, 4.039, 2.160, 5.874, 3.972, 2.000, 2.912, 4.577, 5.728, 2.940]
%min() = 1.582
%max() = 5.874
% med() = 2.912
%compare to platescale ratios 1.59 / 0.6 = 2.65 

The diameters we obtain from the FWHM of the Gaussian fitting from multiple angles allow us to investigate how the diameters change at the same point along the track due to the instrument and line of sight.  \change{The diameters from the fitting were corrected for the using instrumental point spread function \citep[PSF; FWHM of 2.2 EUVI pixels and 2.4 AIA pixels, respectively;][]{2008ApJ...680.1477A,AIA171PSF} and converted to Mm.}  As in the previous section, the regions of good fit were informed by the behavior of the centroid.  These regions of good fit are again between the dotted vertical lines in the left panels of Figures \ref{mod10fits}, \ref{mod13fits}, and \ref{fig:dual}.  For all POIs in the regions of good fit, the AIA diameters were plotted against the corresponding EUVI diameter.  These are seen on the right panels of Figures \ref{mod10fits} and \ref{mod13fits}, and the upper right panel of \figref{fig:dual}.  In general across all tracks in the set, we find no consistent trends in the observed diameter as a function of distance along the track.  However, this could be attributed to the track-to-track variation in what portion was well fit to the data; trends may be obscured due to looking at different portions between tracks.

For these examples and for the other cases in general, we find systematically larger diameters as determined from EUVI.  We define a ratio of median EUVI diameters to that of the AIA diameters in the well fit region.  For Example 1 (\figref{mod10fits}) this value is $\mathchange{4.52}$, and for Example 2 (\figref{mod13fits}) it is $\mathchange{4.59}$.  For all examples, the diameter ratio statistic is in the range $\mathchange{[1.36, 9.18]}$ with a median value $\mathchange{2.20}$.  This \change{median value} is similar to the ratio of the EUVI to AIA platescales, $2.65$.  Thus, though we find the values for EUVI diameters consistently larger than those from AIA, we cannot discount that the difference in the instruments themselves account for this discrepancy.  However, we can propose another interpretation:  Though there is nothing special about the Sun-Earth line, the LOS from AIA aligns with the direction radially outward from the Sun while EUVI's LOS is perpendicular to that.  As the loops (and tracks) have a significant portion mostly in the radial direction, it is possible that the viewing angle does play a role if the loop expands anisotropically with special consideration for the $\hat r$ direction.  Further the sparseness of the intensity information due to the resolution of EUVI with respect to the Gaussian fitting may lead to an overestimation of the diameters in EUVI.  A slight dip in the intensity/radius curve (see \figref{fig:step3}) that may be indicative of two different structures in close proximity on the POS might be ignored by this particular fitting procedure.

\change{Despite these possible shortcomings, we find loop widths in this study similar to others reported in the literature.  The range of median loop diameters was $[0.72, 10.43]$ Mm, with median of all median for diameters $2.07$ and $6.61$ Mm for AIA and EUVI, respectively.  In a study using the same loop catalog and AIA data, \cite{McCarthy2019} found the median of all median diameters was $5.9$ Mm.  However, they noted that their method of boxcar smoothing the intensity profile from which the loop widths were found might lead to an overestimation of diameters and further neglected to account for the instrumental PSF.  In other studies of loop widths with AIA and TRACE \citep[as in][among others]{2000ApJ...541.1059A,2005ApJ...633..499A,2005ApJ...630..596L,2006ApJ...639..459L,2013SoPh..283....5A} the range of widths were comparable to those obtained in this study.  The width distribution from Hi-C reported in \cite{2013ApJ...772L..19B} peaks at $0.272$ Mm with range $[0.09, 0.973]$ Mm, but that instrument has significantly higher spatial resolution.  There are some instances where some of the derived loop diameters are too large (as in \figref{mod10fits} or \ref{fig:dual}).  Due to the limitations in loop selection by both the believability of track-loop matching and quality of the data, some amount of outliers in deriving widths of loops could not be completely eliminated.  However, we sought to insulate our results as best as possible by looking at overall trends instead of individual data points.}

%\change{Indeed, we find diameters larger... in the study of the same AR loops... median value of \cite{McCarthy2019} was 5.9Mm and here it is 2.073
%Longcope 2005 was 3.7 Mm 
%Aschwanden 2000 [1.5Mm to 7.5Mm] Trace, 2013 (loop width distribution [3,9]Mm range peaks at w = 5.10 ± 0.47 Mm (about nine AIA pixels))}

%\change{Hi-C: distribution peaks at 0.272 Mm, range [0.09, 0.973] }
%AIA diameters [2.289, 1.998, 4.952, 4.424, 2.073, 1.844, 1.879, 1.781, 3.253, 0.951, 3.134, 0.719, 2.268] median 2.073
%EUVI diameters [4.067, 3.395, 7.192, 7.432, 9.377, 3.760, 10.425, 8.183, 4.418, 2.655, 6.887, 6.605, 6.152] median 6.605

\change{Though we derive some amount of diameter variation over the length of the loop, our results cannot say anything definitive about loop expansion overall.  Previous studies using SXT \citep{1992PASJ...44L.181K, 2000SoPh..193...53K} and TRACE \citep{2000SoPh..193...77W,2006ApJ...639..459L} have concluded that there is a slight increase in a loop's width at its middle when compared to its ends.  As our regions of good fit tended to include the middle of the loop and exclude the ends, we cannot make a direct comparison.  \changetwo{Considering shorter segments of loops might put a lower limit on loop expansion.}  However, even within our well-fit regions we have changes in loop diameter of up to a factor of $2$.  Though some amount of width fluctuation over short distances has been reported \citep[$\sim25\%$ of the loops average width,][]{2008ApJ...673..586L}, what we find here is much larger.}
%Finds generally slightly larger at the loop's middle compared to the ends.  Our regions of good fit usually gave us the middle, cannot do a direct comoparison.  In regards to diameter fluxuating over the length of the loop }
%\change{\colorbox{cyan}{Compare to similar studies re: (1) diameter} (diameters are large) (abundant transition region moss in the AIA data or from overlapping of structure over the limb in EUVI) and compare to (2) quasi-constant cross sections (SXT, Trace) 
%\cite{2000SoPh..193...77W}: Not significantly thicker at the middle rel to ends, Small expansion factors do not depend on loop length. sim to results of next two lines
%results of Klimchuk+1992 \cite{1992PASJ...44L.181K} loops do not expand
%\cite{2000SoPh..193...53K}: Yohkoh SXT, loops slightly (30\%) wider at middle than feet.  some variability we observe.
%Our results have larger variation over the loop's length same
%\cite{2008ApJ...673..586L}:
%\cite{2008ApJ...673..586L} variation 25 perent over short lengths}
\change{Other studies may make the decision in their selection of loops to choose those that are well isolated from the background.  In this work, the trade off for having our viewing angles almost in quadrature with loops above the limb in our second viewing angle means some compromise for possible  overlap of these EUV features along the LOS.  We also see through more of the solar atmosphere looking on the limb {\em vs.} on the disk.}  \changetwo{This would be a factor in \changethree{seeing variation in loop widths on the limb due to the LOS passing through more material (as in the EUVI angles in this work), in comparison to looking straight down at loops (as in the AIA images).}  The AIA observations do, in fact, have only slight to negligible variation in their widths along the loop.}

We find anti-correlations for the AIA vs EUVI diameter plots in \change{7} of the 13 track/loop pairs.  For this subset, it would be difficult to explain this behavior were the cross sections circular.  Therefore, the anti-correlated behavior opens the possibility that the loops are elliptical \citep{Longcope2020}.  For two viewing angles approximately $90^\circ$ apart, an elongated object would appear larger from one perspective than the other.  Such a thing would produce the anti-correlation that we observe.

%    This is not expected for a loop of circular cross section we would have difficulty seeing an anti-correlation if the loops were perfectly circular.  In this subset, would be difficult to explain if the cross sections were circular.  Therefore opens the possibility that the loops are elliptical (LMM+20) where the perspectives roughly 90* apart see 
%such a thing would produce anti-correlation.

\figref{fig:dual} presents a special case, where the data from STEREO-A, -B and SDO is adequate for simultaneous analysis.  The example analyzed in this figure \change{is the yellow track from Figures~\ref{fig:Aexample} and \ref{fig:Bexample}.  Both track/loop pairs belong to the maybe match category}.  % is in the supplemental figures as {\tt model05}.  
As such, we have the unique ability to compare the diameters not only between AIA and EUVI, but between the two STEREO spacecraft.  The lower right panel of that figure is a scatter plot of the EUVI-A diameter vs EUVI-B.  As the regions of good fit are not exactly aligned for the track in the two different instruments, the gray crosses are the well-fit locations along the track in one instrument while the black crosses are well-fit in both.  The median diameter ratio of A to B is $\mathchange{1.10}$ for all points and $\mathchange{0.88}$ for the black crosses only.  We note that the ratio statistic for all points is larger in part due to more saturation in the A image.  

We find a negative correlation for both black crosses %\change{\colorbox{cyan}{(Spearman's $\rho = -0.65$) 
and the set of all points%($\rho = -0.41$)}}
.  The previous discussion of anti-correlation would not apply if the spacecraft were $180^\circ$ apart.  Though the spacecraft are nearly out of phase, they are not precisely $180^\circ$ apart.  This leaves the possibility open that the slight difference in the respective LOS of the spacecraft and geometry of the coronal loops (i.e., the loop does not have an isotropic, cylindrical cross section) lead to the difference in diameters.  Though this is one explanation for the anti-correlation between the diameters obtained from the two STEREO vantage points, there may be other mechanisms at work.

\subsection{Diameter and intensity comparison}\label{sec:ac}

\begin{figure}[tbhp]
\centering
     \includegraphics[width=0.999\linewidth]{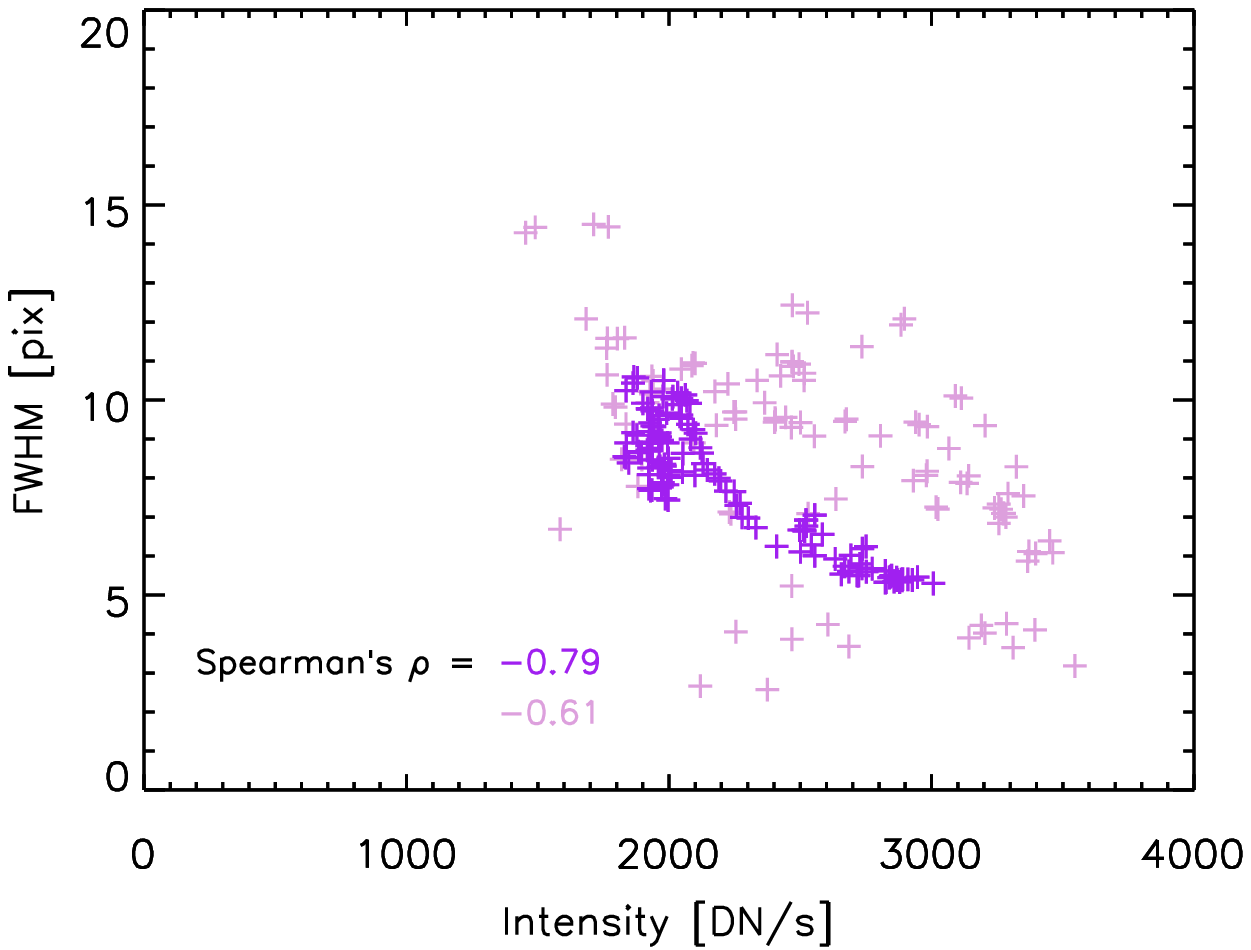}
     \caption{A scatter plot showing the relationship between the intensity and width of the loop at all POIs that define the track in EUVI for Example 1, for which there is a negative correlation.  The dark purple crosses are the points in the well fit region with Spearman's $\rho = -0.79$ $\mathchange{(p = 7.4 \times 10^{-32})}$.  All fits along the track include the light purple data points and for the entire plot $\rho=-0.61$ $\mathchange{(p = 9.0 \times 10^{-28})}$.}
     \label{fig:anti}
\end{figure}

Additional evidence of elliptical cross sections can be found by comparing loop diameter to brightness.  For a loop expanding non-isotropically along its axis but with a constant filling factor, we would expect narrower sections to be bright and for wider sections to be \change{dimmer}.  That is, if we view one loop along the direction of its major axis our LOS passes through more emitting material in the optically thin corona, and appears brighter than if viewed along the minor axis.  To elucidate this hypothesis, we examined how our Gaussian fitting parameters $A$ (Gaussian peak, used as a proxy for intensity) and FWHM compared.  A plot of this behavior for Example 1 as fit from the EUVI data is shown in \figref{fig:anti}.  Though this is the most dramatic example of this type of behavior, a significant fraction \change{of loops} exhibit an anti-correlation between FWHM diameter and intensity for EUVI.  That being said, we note that Example 1 does not exhibit the anti-correlated behavior demonstrated in the diameter-diameter plot in right panel of \figref{mod10fits}.  

\change{Our result seems to contradict some previously reported.  Analysis of loops observed in $193${\AA} with Hi-C by  \cite{2020ApJ...900..167K} found either uncorrelated or positively correlated behavior of intensity vs widths.  They cited these results as evidence of approximately circular cross sections in the examined loops, but did not extend this inference to loops universally.  The results we present in this section are the inverse of those results; we find our intensity and width measurements are more likely to be uncorrelated or inversely related.  }%The differences may be attributed to partitioning -- which itself would select different populations of loops by the nature of different temperature of passband -- or by the selction of loops not comperable in size.}\footnote{  \colorbox{cyan}{Did they look at shorter loops?}  Compare, histogram of length and height}

\begin{figure}[tbhp]
\centering
     \includegraphics[width=0.999\linewidth]{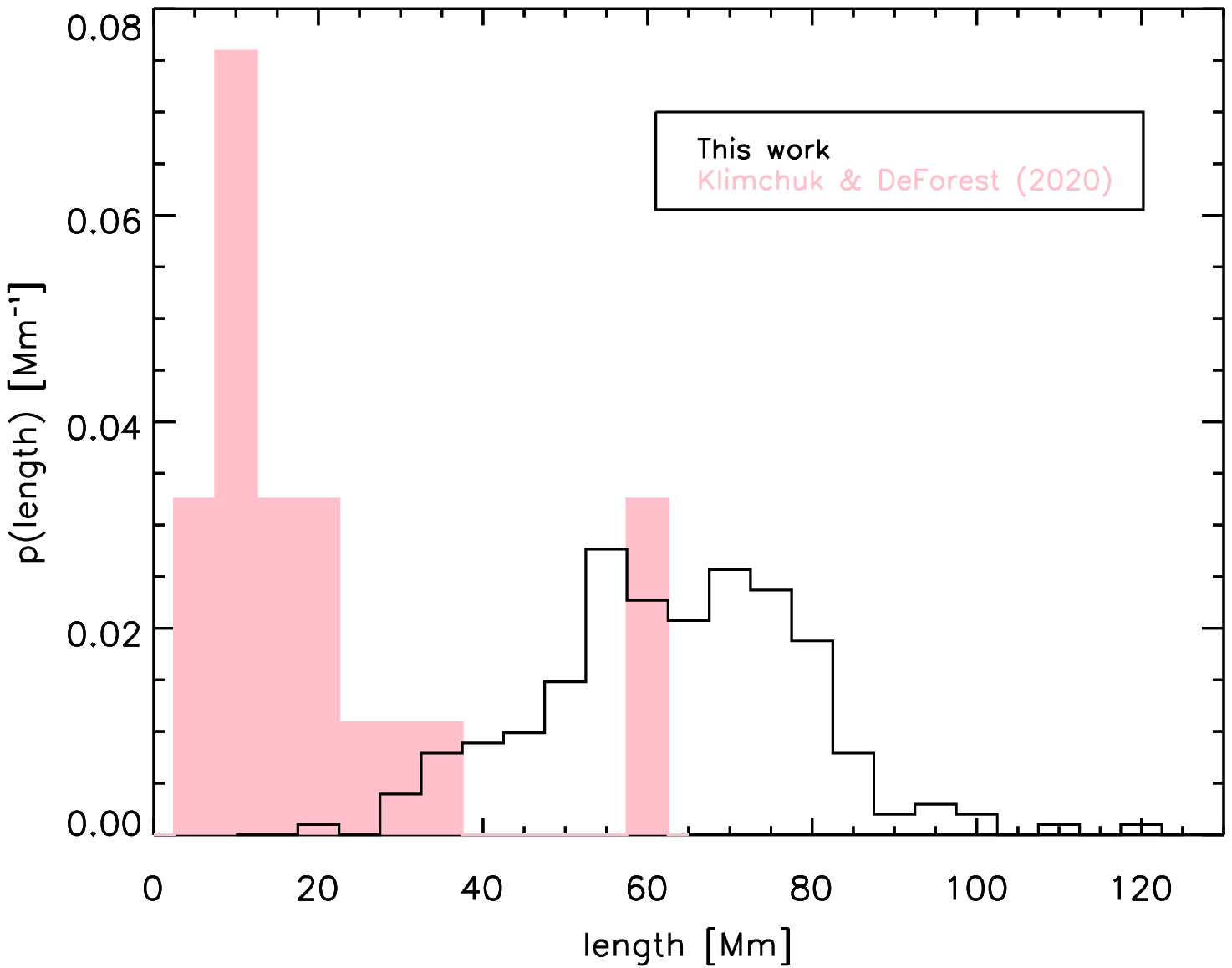}
     \caption{\change{A histogram comparing the lengths of loops observed in this study (black) to those of \citet[][pink, see Table 1 in that work]{2020ApJ...900..167K}.  These data have been normalized such that the y-axis shows the value of the probability density function.  \citet{2020ApJ...900..167K} in general examined shorter loops \changetwo{(22 loops total)} than those herein \changetwo{(151 loops)}.}}
     \label{fig:lengths}
\end{figure}

\change{This discrepancy may be attributed to observational wavelengths (which itself would select different populations of loops on the basis of temperature) or by the nature of the loops of interest themselves.  Observations in $193${\AA} (log(T) = 6.2 K) pick up hotter plasma than those taken in $171${\AA} \citep[log(T) = 5.8 K;][]{2012SoPh..275...17L}.  Further difference in these data sets might be attributed to the sizes of the studied loops.  The length differences between the populations can be seen in the histograms plotted in \figref{fig:lengths}, where black shows loops from this work and pink show those from \citet[][pink; see Table 1 in that work]{2020ApJ...900..167K}.  Generally, \citet{2020ApJ...900..167K} examined shorter loops when compared to this study.  Both the Hi-C resolution was finer and the FOV was smaller than AIA or EUVI.  }

In the respective well fit regions of the 13 track/loop pairs viewed in EUVI, 6 had a negative value for $\rho$, and 4 had $\rho < -0.4$.  Those cases with $\rho > 0$ covered the range $[0.007, 0.870]$, but tended to cluster around the range's minimum and maximum values.  The instances of fitting from AIA do not exhibit such behavior consistently.  \changetwo{The results of the diameter/intensity correlation were scattered enough such that we could not be conclusive about the trends observed with this spacecraft for these data.}

\change{The lack of an anti-correlation in the AIA intensity vs diameter on its own would not be evidence of anisotropic cross sections.  When combined with the anti-correlated behavior of the EUVI observations, we see changing diameters from one perspective but not from the other.  This might be further evidence of anisotropic expansion or some other mechanism in which a changing aspect ratio is observed.}

%\change{%\colorbox{cyan}{Compare to Hi-C here?} 

%However, if this interpretation werecorrect, the brightflux tubes would not have circular crosssections, except perhaps at a single location in one or both legs.We do not wish to suggest that our inference of circular crosssections is definitive or applies universally to all loops. It seemslikely that the Malanushenko & Schrijver interpretation iscorrect for many loops, but what fraction of loops is an openquestion.

\section{Conclusions and outlook}\label{sec:conclusions}

One of the primary aims of this study was the usage of a second perspective to confirm the veracity of the $\alpha$-h fitting method.  \change{Of the 151 tracks that were viable for visual comparison, 40 (26.5\%) were excellent match, 65 (43\%) were maybe match, and 46 (30.5\%) were no match.}  Overall, we find good agreement in the co-alignment of tracks to loops observed in EUVI; approximately $70\%$ of our tracks had some feature to which we could align them.  This finding lends a novel line of support in favor of the fidelity of the $\alpha$-h fitting method.  As to the remaining $30\%$ that did not seem to correspond to any features in the corona, to first order we might expect these ``missing'' features to indicate that the $\alpha$-h fitting went awry in the construction of tracks.  

However, this conclusion relies on the assumption that loops in the corona are monolithic objects, existing and observable from varying vantage points.  We cannot discount the possibility that the track is an accurate fit, but due to factors like anisotropic expansion we do not see enough of a density enhancement in the alternate LOS for the loop to be visible.

%\change{We also contend that four of the loops observed within this study \changetwo{are consistent with} elliptical cross sections, as defined from having a negative correlation in the behavior in both the diameter-diameter and diameter-intensity comparisons.}
\change{We also contend that four \changetwo{of the 13 loops deemed suitable for the quantitative analysis of their diameters} within this study \changetwo{are consistent with} elliptical cross sections, as defined from having a negative correlation in the behavior in both the diameter-diameter and diameter-intensity comparisons.  \changetwo{Of the remaining nine loops, four exhibited correlated behavior in both comparisons, and five had a combination of correlated and anti-correlated behavior.}}

The results of \secref{sec:diam} point to systematically larger diameters observed from EUVI compared to those viewed from AIA.  An optimist would point to this as evidence of smaller diameters from AIA compared to EUVI, similar to the idea from \citet{2013ApJ...775..120M} of preferential loop expansion along the AIA LOS for this case.  We note the AIA LOS (and Sun-Earth line) is itself not particularly special, however it corresponds to the $\hat r$ direction and both EUVI LOS are perpendicular to that.  This could be investigated in the future by looking at loops that lie at disk center in EUVI and on the limb from AIA.  A more likely explanation, however, is instrumental effects between EUVI and AIA playing the more influential role in the diameter differences.  This is evidenced by the median diameter ratio statistic for all tracks we analyzed quantitatively being comparable to the ratio of the two different instrument's plate scales.  Further, the coarser resolution of the EUVI instrument combined with the overlapping of loops due to the above-the-limb orientation of the region of interest may have contributed to an overestimation of loop diameters that our methodology could not remedy.  A small dip in the intensity/radius profile might indicate a ``boundary'' between two overlapping flux tubes, but might be ignored by our straightforward Gaussian fitting.  Improving on the techniques described in this study may remedy some of the issues described above, particularly, for example, by the application of a machine learning or Markov Chain Monte Carlo method.  These more complex techniques could, in theory, compensate for the sparsely gridded information and overlapping of fine, sub-resolution strands of flux.

We find a fair amount of anti-correlation (determined by a negative Spearman's $\rho$) between EUVI and AIA diameters.  One might expect some amount of correlation between the two properties if we were truly looking at a monolithic object, even for diameters consistently larger in one instrument than the other.  That is, it might be expected that a loop's major and minor axes scale proportionally.  However, this has not been the case for the track/loop pairs analyzed in this study.  

One possible explanation is a type of mechanism wherein, during loop expansion, the major and minor axes change in an inversely proportional manner.  This type of relationship between the axes is one that would would yield the anti-correlated diameter trend we have seen, and point to a changing aspect ratio along these loops.  It is also possible that the loop's cross section has a constant aspect ratio with the ellipse rotating about its center (e.g., as in a twisting ribbon).  Either idea is bolstered by the findings of anti-correlated behavior between the intensity (Gaussian peak) and FWHM of our fitting procedure as detailed in \secref{sec:ac}.  \change{This is despite the conflicting results in the intensity-diameter comparison between this work and that of \cite{2020ApJ...900..167K}, which may speak to different populations of loops existing in the solar atmosphere.  Indeed, the data differs between the two studies both in observational wavelength ($193${\AA} vs $171${\AA}) and length.  }
%\footnote{\colorbox{cyan}{193 vs 171 so they are different populations in terms of temperature.}  Also their AR was very complicated mess in the core where they looked}}  
Our results thus appear consistent with the hypothesis that some loops are monolithic objects with elliptical cross section, though we note that this is but one proposed explanation.  

Nevertheless, this poses new questions for what physical process might lead to this behavior, if such behavior is even a real property of a structure that exists, or if we are -- like in the traditional STEREO \change{stereoscopy} -- identifying coinciding features merely because they are there.  Perhaps the structure of loops in the corona are even more different than we might expect from the common conceptions in the field.

\section{Acknowledgments} 
This work was supported by NASA's HGI program.

\change{We would like to extend our thanks to the anonymous referees for the valuable input in regards to the improvement of this work.}

\bibliography{paperbib}
\bibliographystyle{aasjournal}

\end{document}

%% file: figset.tex
%\figsetstart

\begin{figure}
\plotone{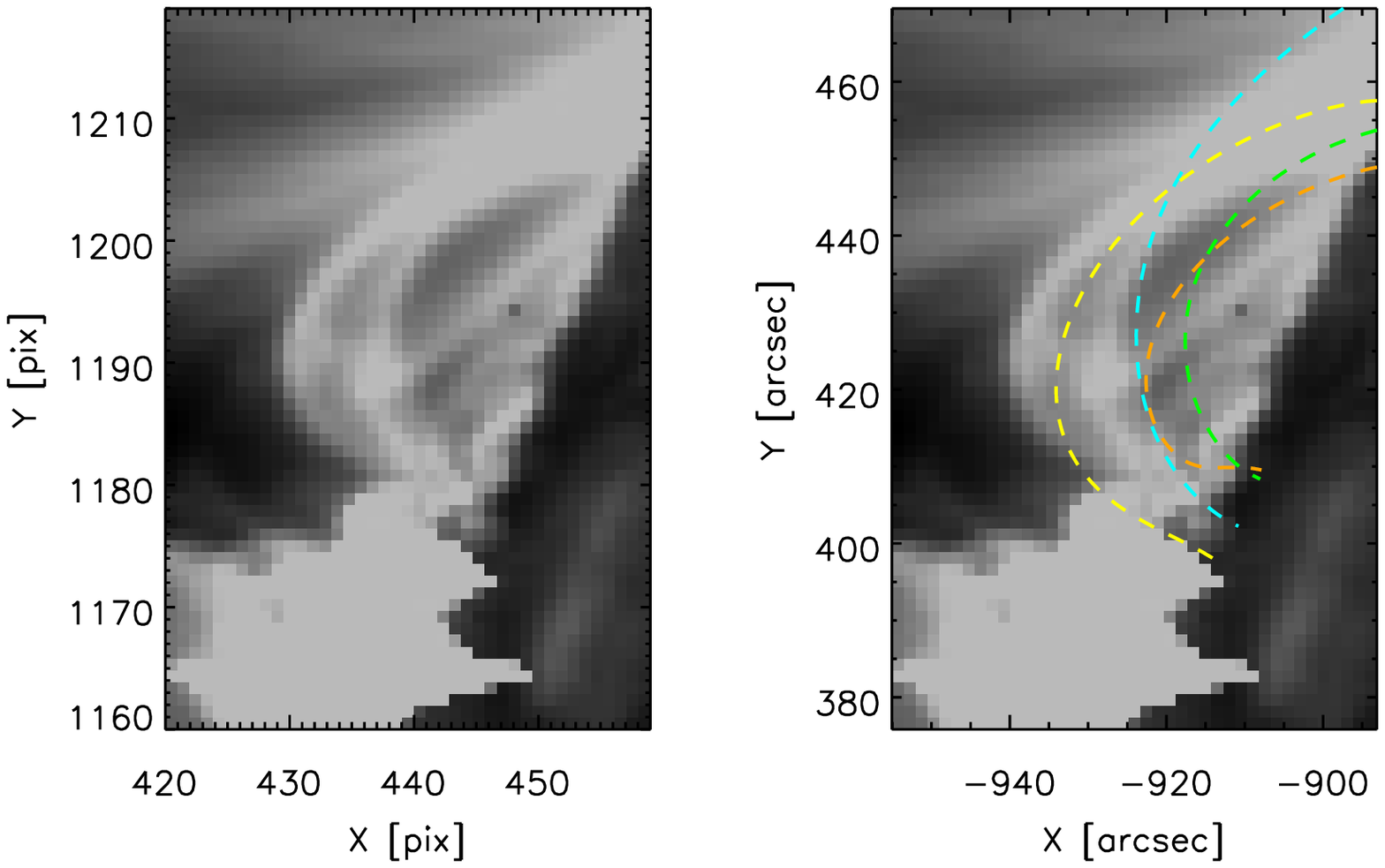}
\caption{\change{An image taken from EUVI-A $171${\AA} with the corresponding tracks plotted over.  The complete figure set (18 images) is available in the online journal.}}
\label{fig:Aexample}
\end{figure}

\begin{figure}
\plotone{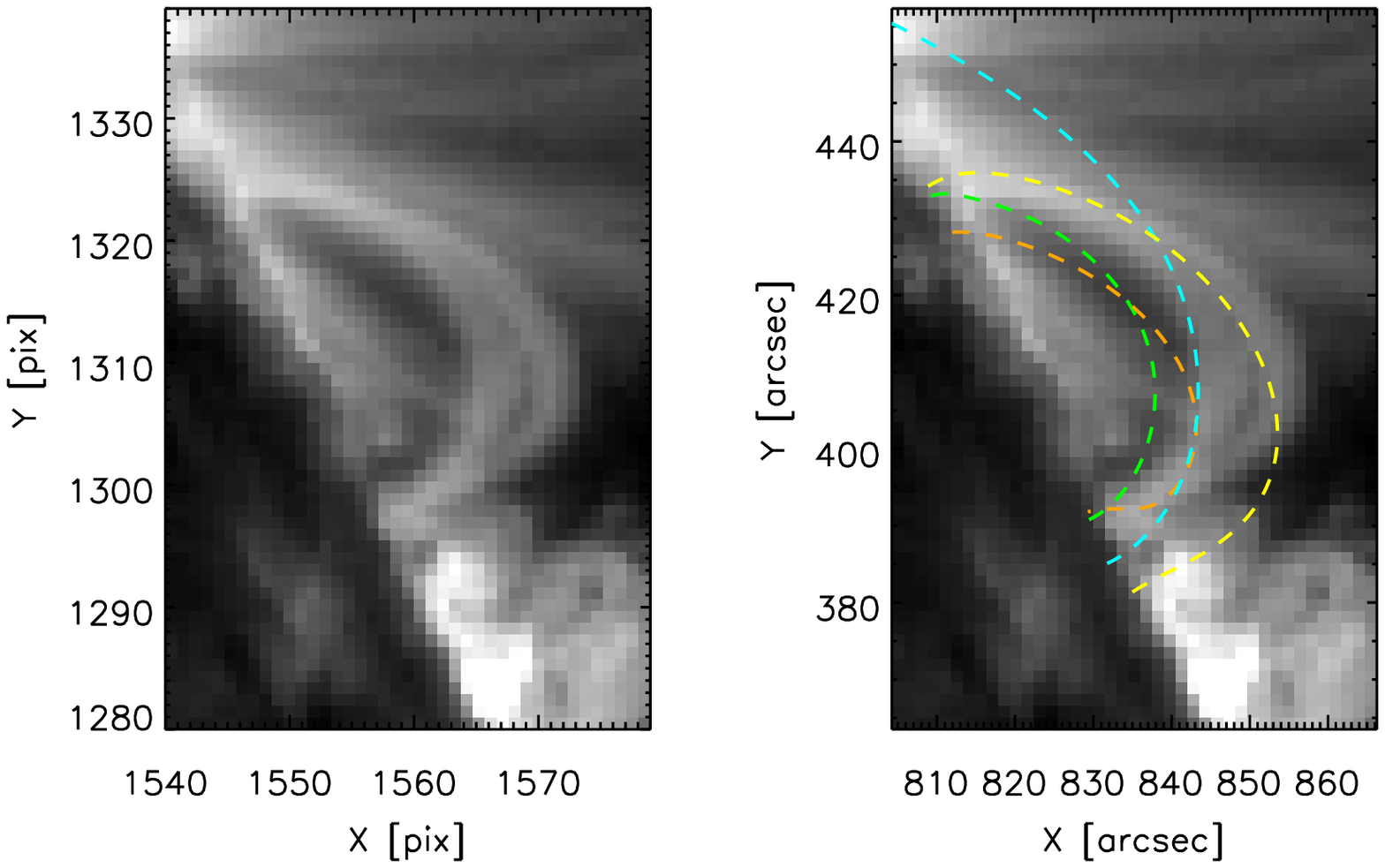}
\caption{\change{An image taken from EUVI-B $171${\AA} with the corresponding tracks plotted over.  The complete figure set (26 images) is available in the online journal.}}
\label{fig:Bexample}
\end{figure}